\documentclass[acmlarge,screen]{acmart}

\usepackage{graphicx}
\usepackage{longtable}
\usepackage{multirow,colortbl}
\usepackage{titlesec}
\usepackage{url}
\usepackage{adjustbox}

\def\BibTeX{{\rm B\kern-.05em{\sc i\kern-.025em b}\kern-.08emT\kern-.1667em\lower.7ex\hbox{E}\kern-.125emX}}

\begin{document}

%
\title{Methods and advancement of content-based fashion image retrieval: A Review}

%
\author{Amin ~Muhammad Shoib}
\affiliation{%
	\institution{School of Software Engineering, East China Normal University}
	\streetaddress{3663 North Zhongshan Road.}
	\city{Shanghai}
	\country{China}}
\email{52184501030@stu.ecnu.edu.cn}

\author{Jabeen Summaira}
\email{11821129@zju.edu.cn}
\affiliation{%
	\institution{College of Computer Science and Technology, Zhejiang University, China}
	\streetaddress{P.O. Box W-99}
	\city{Hangzhou}
	\postcode{310027}
}

\author{Changbo Wang}
\affiliation{%
	\institution{School of Computer Science and Technology, East China Normal University}
	\streetaddress{3663 North Zhongshan Road.}
	\city{Shanghai}
	\country{China}}
\email{cbwang@cs.ecnu.edu.cn}

\author{Abdul Jabbar}
\email{Jabbar@zju.edu.cn}
\affiliation{%
	\institution{College of Computer Science and Technology, Zhejiang University, China}
	\streetaddress{P.O. Box W-99}
	\city{Hangzhou}
	\postcode{310027}
}

%
\renewcommand{\shortauthors}{Amin Muhammad Shoib et al.}

%
\begin{abstract}
Content-based fashion image retrieval (CBFIR) has been widely used in our daily life for searching fashion images or items from online platforms. In e-commerce purchasing, the CBFIR system can retrieve fashion items or products with the same or comparable features when a consumer uploads a reference image, image with text, sketch or visual stream from their daily life. This lowers the CBFIR system's reliance on text and allows for a more accurate and direct searching of the desired fashion product. Considering recent developments, CBFIR still has limits when it comes to visual searching in the real world due to the simultaneous availability of multiple fashion items, occlusion of fashion products, and shape deformation. This paper focuses on CBFIR methods with the guidance of images, images with text, sketches, and videos. Accordingly, we categorized CBFIR methods into four main categories, i.e., (i) image-guided CBFIR (with the addition of attributes and styles), (ii) image \& text-guided, (iii) sketch-guided, and (iv) video-guided CBFIR methods. The baseline methodologies have been thoroughly analyzed, and the most recent developments in CBFIR over the past six years (2017–2022) have been thoroughly examined. Finally, key issues are highlighted for CBFIR with promising directions for future research.
\end{abstract}

%
%


%

%

%
\maketitle

 \section{Introduction}
 \label{intro}
 
 With technological advancements, the fashion industry has regained its imperativeness since combining with industries like multimedia, computer systems, and the world wide web. In the latest decades, fashion study has achieved substantial improvements, with the utilization of source images for fashion retrieval becoming one of the most successful methods and also a study hotspot \cite{li2021recent,mo2022concentrated}. Daily countless images and their related digital data are added to e-commerce web pages, exacerbating the fashion retrieval issue. The introduction of technological advances has changed the working trend of nearly all vital activities, as well as internet shopping is one such area in which it is causing significant chaos \cite{rahman2018consumer}. To offer a genuine and satisfactory online purchasing experience, personalized focus to every customer is a requirement which is also a significant challenging issue for the fashion industry.
 
\begin{figure*}
 	\caption{General structure diagram of CBFIR models.}
 	\centering
 	\includegraphics[width=\textwidth]{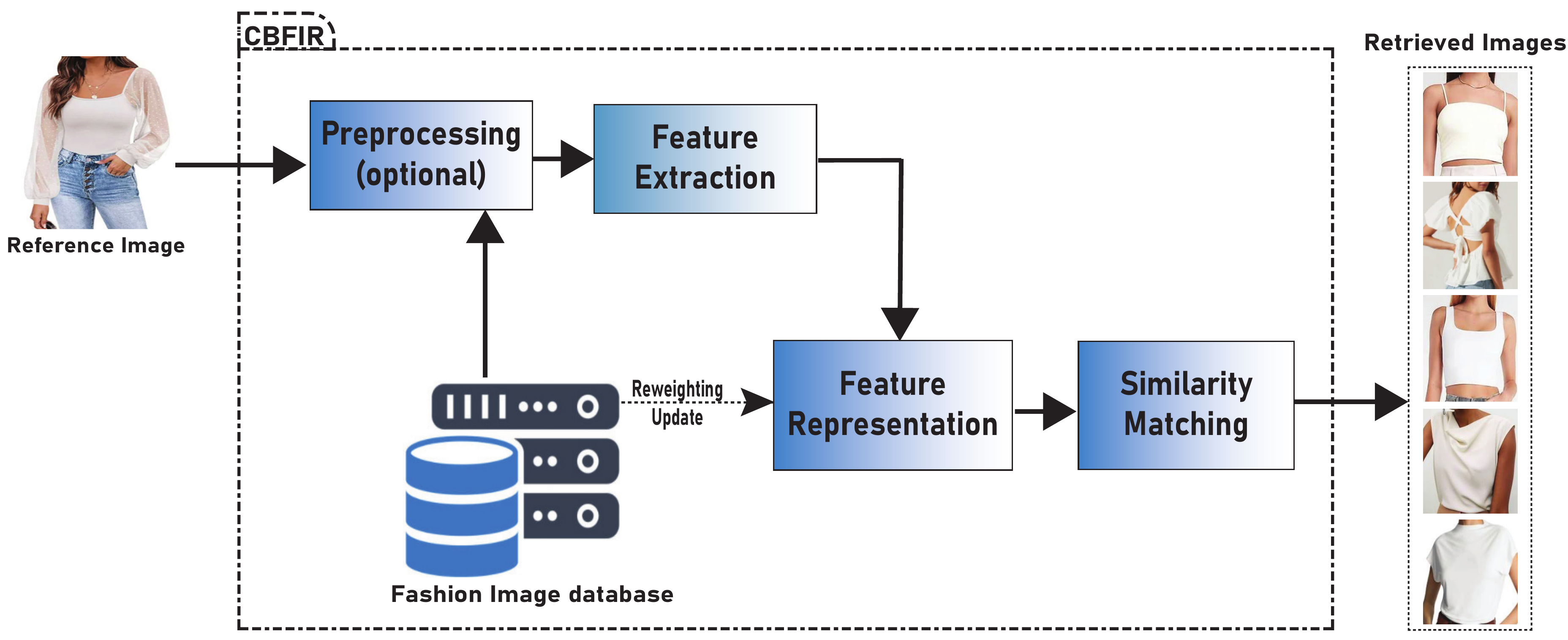}
 	\label{fig:gStructureCBFIR}
 \end{figure*}

 The goal of content-based fashion image retrieval (CBFIR) is to find fashion items or products from a database of photographs that are aesthetically the most comparable to a particular querying image \cite{nodari2012mobile}. The general structure diagram of CBFIR models illustrates the procedure in Figure \ref{fig:gStructureCBFIR}. Approaches including classification and regression, generative adversarial networks, clustering algorithms, natural language processing, and computer vision are currently at the edge. Computer vision is mostly used in the fashion industry to analyze photographs, identify objects, and identify appropriate imagery for fashion products recommendations  \cite{ain2017sentiment,cheng2021fashion,liu2018deep,alamsyah2019object,suekane2022personalized}. For everyday particular fashion photos, several experts have published their concepts for efficient extracting features and precise attribute/object recognition AI-based methods. For everyday particular fashion photos, several experts have published their concepts for efficient extracting features and precise attribute/object recognition Intelligence methods \cite{liu2014fashion,park2019study,kim2021self,amin2022fashion}. These methods are quite good at helping you understand what consumers are expressing about the things they are purchasing, which can be used as a foundation for upcoming business plans.

 CBFIR is frequently employed in e-commerce systems and search areas like Google, Taobao, Baidu, Jingdong,  and many others to retrieve the desired fashion images. Individuals enjoy taking images of their everyday surroundings and shopping online for their favourite items. Customers may rapidly and precisely locate related apparel styles online with the use of CBFIR technology \cite{nodari2012mobile,eswaran2022reverse,islam2021ornament}. This not just satisfies everyday requirements and raises the standard of lifestyle for individuals, but it also encourages the purchase of apparel. The size of the local and international fashion marketplaces is consistently expanding, in accordance with the research of the industry. Despite technology for retrieving images of clothes having advanced significantly over the previous two decades, clothing image retrieval in cross-domain scenarios still presents significant hurdles \cite{shah2020random,chen2018generic,xiang2019fabric}.
 
 In this paper, we review the different techniques used for the CBFIR task, and we categorize these recent techniques into four major groups. This grouping will help researchers in the future to find out the research problem in the fashion retrieval domain. The fashion image retrieval task is mainly based on content provided to the system to find similar products. According to recent advancements in CBFIR, we divided the domain area into four major groups, i.e., image-guided CBFIR methods, image + text-guided CBFIR methods, sketch-guided CBFIR methods, and video-guided CBFIR methods.

 \subsection{Relevance to other review articles and Contribution}
 
 In order to bring our survey into perspective, we found and provided a relevant review and surveyed articles to explain how our paper varies from and builds upon prior research. Recently, various surveys were already published relating to the subject topic of fashion image retrieval. 
 Liu, S. et al. \cite{liu2014fashion} discussed a brief discussion about fashion analysis techniques (including; clothing modelling, clothing recognition, clothing parsing, clothing retrieval, and clothing recommendation). Significantly less literature is provided in this article regarding the clothing retrieval part. 
 Cheng, W.H. et al. \cite{cheng2021fashion} provides a comprehensive survey on fashion-related articles. The author categorized the article into four main categories, i.e., fashion detection, fashion analysis, fashion synthesis, and fashion recommendation. In this survey, the author describes a small portion of fashion item retrieval and only discusses a few FIR methods from 2011 to 2018. 
 Ning, C. et al \cite{ning2022survey} survey article is very close to our survey, which discussed cross-modal fashion image retrieval models with the perspective of deep metric learning and critical region recognition. 
 
 Recent studies reveal that there are two key challenges with it: (1) Clothing is a flexible material, so when seen from varied shooting positions or When wearing various body kinds, the look might be highly diverse. The user-provided query image could have been taken under difficult situations, with a challenging background, different filming angles, different brightness situations, or perhaps occlusion. However, the majority of shop photographs have tidy backdrops, excellent lighting, and favorable frontal perspectives. (2) It is a natural property of clothing picture data that the intra-class variation is high and the inter-class variance is low. For instance, two dresses from other categories seem to be very identical in terms of style and color, but the necklines of one dress are V-shaped while the other is U-shaped, which is a little noticeable variation. A garment with a U-neck would not be regarded as a suitable search result in the retrieval system provided a picture of a user's clothes with a V-neck.
 However, we categorize the whole domain of CBFIR in a distinctive way that is helpful for researchers in the future with this analysis. Our key contributions are distinctive in that way; we aim to review the research from contemporary advances in fashion image retrieval according to the presented categories. The following list summarizes the important contributions of our article:

 \begin{itemize}
 	\item We provide a brand-new, fine-grained taxonomy of several CBFIR approaches that further detail various groupings of content-based fashion image retrieval tasks.
 	\item We categorized all the techniques or methods of CBFIR into four main categories, i.e., image-guided fashion image retrieval methods, image \& text-guided methods, sketch-guided methods, and video-guided methods.
 	\item We outline potential future research directions for CBFIR approaches while highlighting their limitations and issues.
 	
 \end{itemize}

 \subsection{Structure of article}
 The remaining article is structured as follows. 
 The fashion image retrieval's background is discussed in section \ref{R-Work}. 
 Recent methodologies of CBFIR are categorized in section \ref{CBFIR-methods}. 
 Datasets of image-guided CBFIR, image \& text-guided CBFIR, sketch-guided CBFIR, and video-guided CBFIR are presented in section \ref{CBFIR-datasets}. 
 A brief discussion and future research directions are proposed for CBFIR models in section \ref{Discussion-comparativeAnalysis}. 
 Lastly, section \ref{concl} concludes the review article.

 \section{Background}
 \label{R-Work}

 \subsection{Fashion Image Retrieval (FIR)}
 
 \subsubsection{Uni-modal FIR}
\begin{figure}
 	\caption{Unimodal FIR method using image or text as a query.}
 	\centering
 	\includegraphics[width=\textwidth]{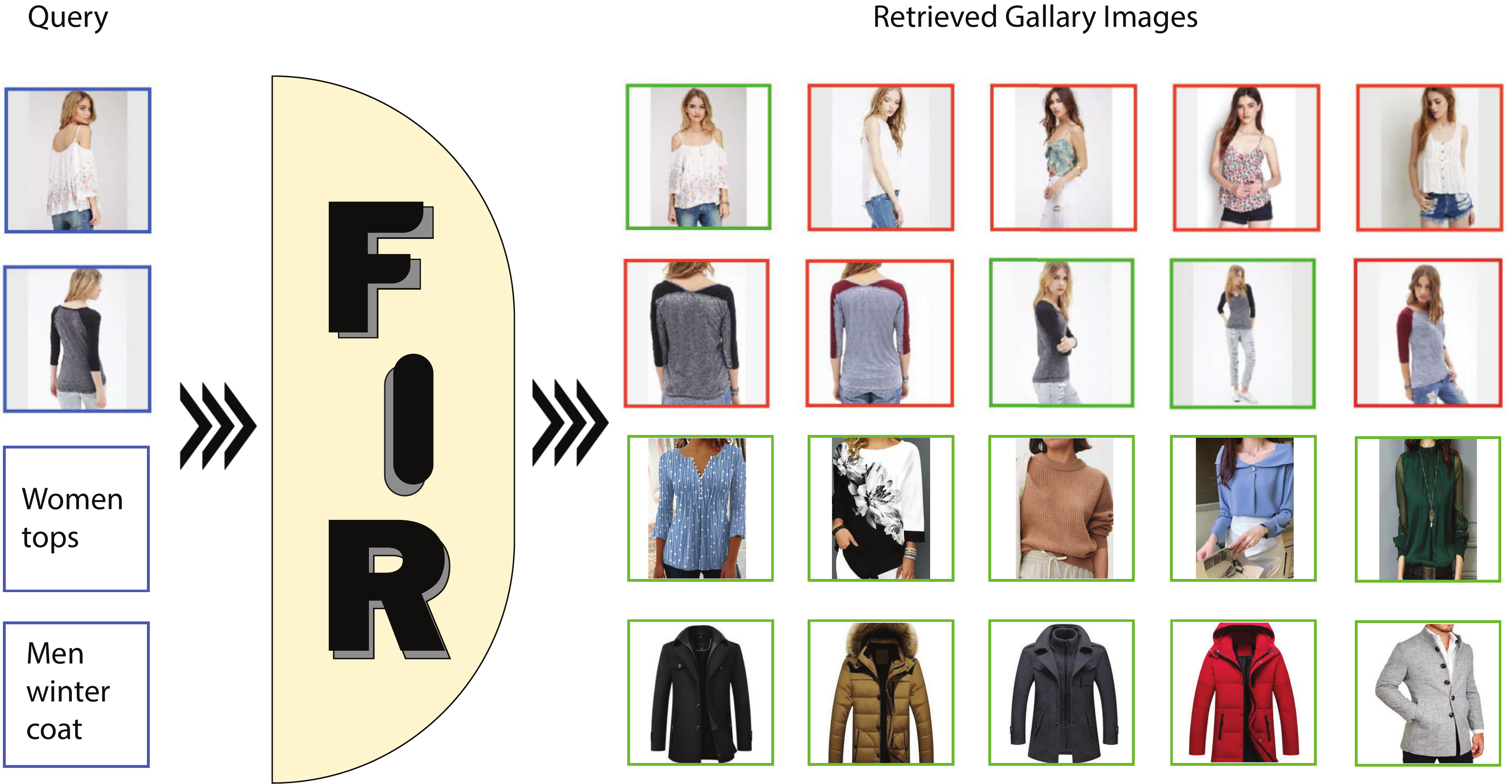}
 	\label{fig:unimodal}
 \end{figure}

 Unimodal FIR methods are used to retrieve desired fashion products or items using a single modality as input to the system \cite{anwaar2021compositional}. This single modality is mostly in the form of text or reference image as a query to the model to retrieve the desired fashion images. Figure \ref{fig:unimodal} shows the general structure of unimodal fashion image retrieval outputs with the query image or text perspective. 
 
 \textbf{Unimodal-Text retrieval systems} use the actual text that already exists on web data, making indexing and keyword extraction rather simple. The architecture of keywords-based or text-based FIR systems is rather simpler compared to content-based FIR. The unimodal-text retrieval models vary from a "simple frequency-of-occurrence-based scheme to the ontology-based scheme." Text-based FIR systems can take care of semantic queries more efficiently compared to content-based FIR models \cite{virmani2019text,riad2012literature,sarwar2013ontology,vijayarajan2012review,wang2013personal}. However, the retrieval process of unimodal-image retrieval systems is distinct from conventional information retrieval models in that image collections are basically unorganized since digital images are nothing more than arrays of pixel intensities without any underlying context. Extracting useful or meaningful information (such as identifying the clothing textures, colors, shapes, etc.) from the input data is one of the important aspects of any fashion image retrieval system before the reasoning stage \cite{chen2012describing,vittayakorn2015runway,yang2011real,tong2016differential,hidayati2017learning,yang2020towards}.
 Earlier unimodal-text retrieval models are mostly centered on manual image annotations, in which the user annotates the content of the fashion products with some keywords. After that, these manual annotations are used to retrieve fashion items or products. The process of manually annotating the image content is subjective and time-consuming. Different annotators annotate the same clothing product with different descriptions. Similarly, one person can make multiple references to the same visual at different times depending on the circumstances. Therefore, manual annotations may be utilized in the context of some specific areas, like digital libraries, virtual museums, personal albums, etc. Automatic image indexing is a possible solution to deal with this issue in text-based retrieval models. There are many techniques for automatic indexing, out of which the most popular one is "count the frequency of occurrence of words." The weighting scheme can be done according to the physical distance of the words for descriptions of an image in the database. And feedback from the consumers about the image results can be used for keyword refinement. The weight assignment procedure to image keywords also relies highly on the domain or context, which is then used for retrieving the desired image \cite{ahamd2003old,mezaris2003ontology,setchi2011semantic,bloehdorn2005semantic,laenen2017cross}. Numerous aesthetic characteristics of fashion products are difficult to describe in words. Text retrieval searches for instances of search terms in product or text descriptions using more recent implicit semantic algorithms.
 
 \textbf{Unimodal-Image FIR} models search and retrieve the same type of fashion products from the large image databases according to the reference input image. Feature extraction techniques play an important part in unimodal fashion image retrieval systems for retrieving similar kinds of fashion products or items from the database \cite{deng2018learning,jiang2018deep,kherfi2004image,hsu2011clothing,mizuochi2014clothing}. Similarity and feature vectors (FV) are two important mechanisms of FIR models. The FV is extracted to signify each reference and database image in the FIR system. The database and reference image similarity is calculated according to the extracted FVs \cite{lin2015rapid,liao2018interpretable,ay2021visual,zhang2020clothingout,he2017fast}. In recent years, researchers contributed to improving the accuracy of unimodal FIR using similarity measurement and feature extraction perspectives. Many artificial intelligence frameworks have been proposed during the last decades to increase FIR performance with the perspective of optimum utilization of computational resources. 
 In unimodal FIR systems, using a reference image as an input allows consumers to transform rich information about the desired fashion product or item as compared to using text as a reference query. Moreover, another benefit of using an image as a query over using text as input is that the language of images is universal to the system \cite{rangkuti2021improved,eshwar2016apparel,han2017automatic,hadi2015buy}. If the user wants some additional attributes about the desired product, then unimodal-image FIR systems cannot retrieve the desired fashion product with additional attributes.
 
 \begin{figure}
 	\caption{Multimodal FIR method using a reference image and textual user requirement as a query.}
 	\centering
 	\includegraphics[width=\textwidth]{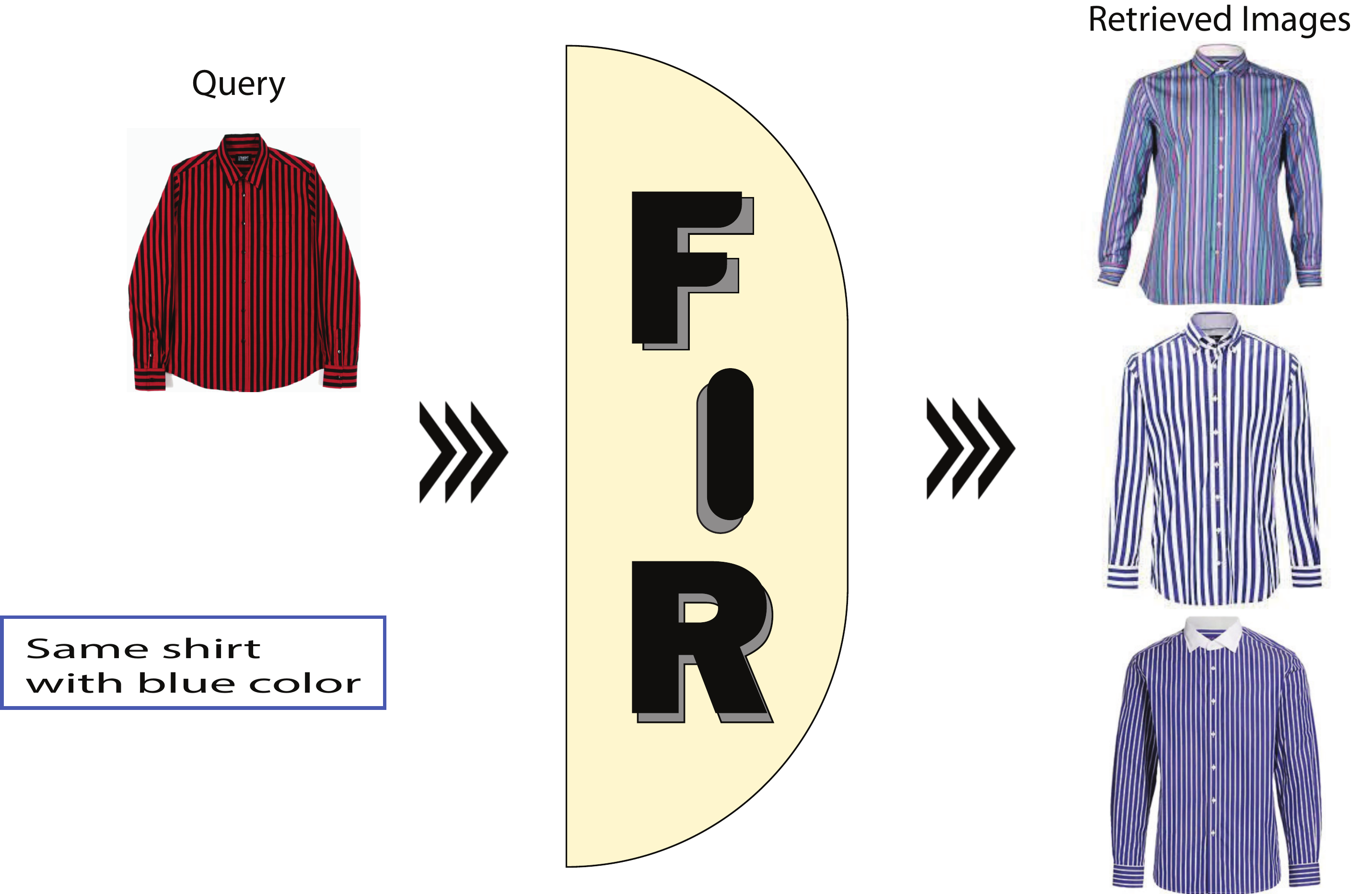}
 	\label{fig:multimodal}
 \end{figure}
 
 \subsubsection{Multimodal FIR}
 
 
 In multimodal FIR systems, more than one modality is used to query the system for fashion image retrieval tasks \cite{jabeen2022review}. Mostly, image, video, and text modalities are fed as input to the FIR systems to retrieve similar fashion images. In multimodal FIR systems, a user can retrieve the desired fashion product by querying the reference image and also fed text to retrieve some additional attributes along with the reference image \cite{miao2020clothingnet,tian2023fashion,jo2020flexible,tautkute2019deepstyle,laenen2018web}. Therefore, multimodal FIR models allow users to retrieve fashion products with images and manipulate the properties of the image with the help of text. The search mechanism of fashion products in cross-modal retrieval systems is done with the intermodal representations for textual and image/video fashion clothing attributes\cite{arslan2019multimodal}. Fashion image or product retrieval with multimodal queries like text or image/video is an innovative search pattern that received great attention from researchers until now \cite{islam2021ornament,chen2017query}. The prevalence of multimodal fashion product data and current developments in deep learning models to analyze linguistic and visual data have increased the popularity of multimodal techniques outside of fashion product search. More specifically, numerous image-text alignment techniques were designed where the recurrent neural network (RNN) generates the image descriptions. Figure \ref{fig:multimodal} presents the general scheme of multimodal FIR systems, where a reference image is queried into the FIR system with the attributes/properties of the clothing product of the user in the form of text to retrieve the desired fashion item or product.

 Given the fact that the internet is expanding quickly, consumers still have difficulty obtaining the necessary and pertinent information despite having access to huge volumes of data. For multimodal fashion image retrieval, this data may be in the form of visual or textual features \cite{liu2019neural,yamaguchi2014retrieving,hsiao2021culture}. Although the abundance of photos available on the internet, image retrieval search engines are still uncommon. Therefore, designing a framework that assists users in finding the desired fashion items or products with good retrieval accuracy and these frameworks can be used for numerous applications in the fashion industry \cite{vanderploeg2017application,unay2012innovation,akram2022implementation,celikik2022reusable,gazzola2020trends,mohiuddin2022role,behr2018fashion}. Improved visualization and comprehension need to learn a common hidden feature representation from different types of clothing data. Fashion clothing products or items made up of various media formats and coming from various resources can be mapped into a uniform latent area using this unifying latent feature representation, where metric distance measurements are used for fashion analysis, recommendations, retrieval, and some other applications \cite{gu2018multi}. Fashion items or products often comprise images and related properties normally in the form of text like titles, tags, sized, short descriptions, colors, etc. In multimodal FIR systems, these reference clothing images and their properties play an important part in retrieving the desired items or products.
 
 \begin{figure}
 	\caption{Taxonomy diagram of content-based fashion image retrieval.}
 	\centering
 	\includegraphics[width=\textwidth]{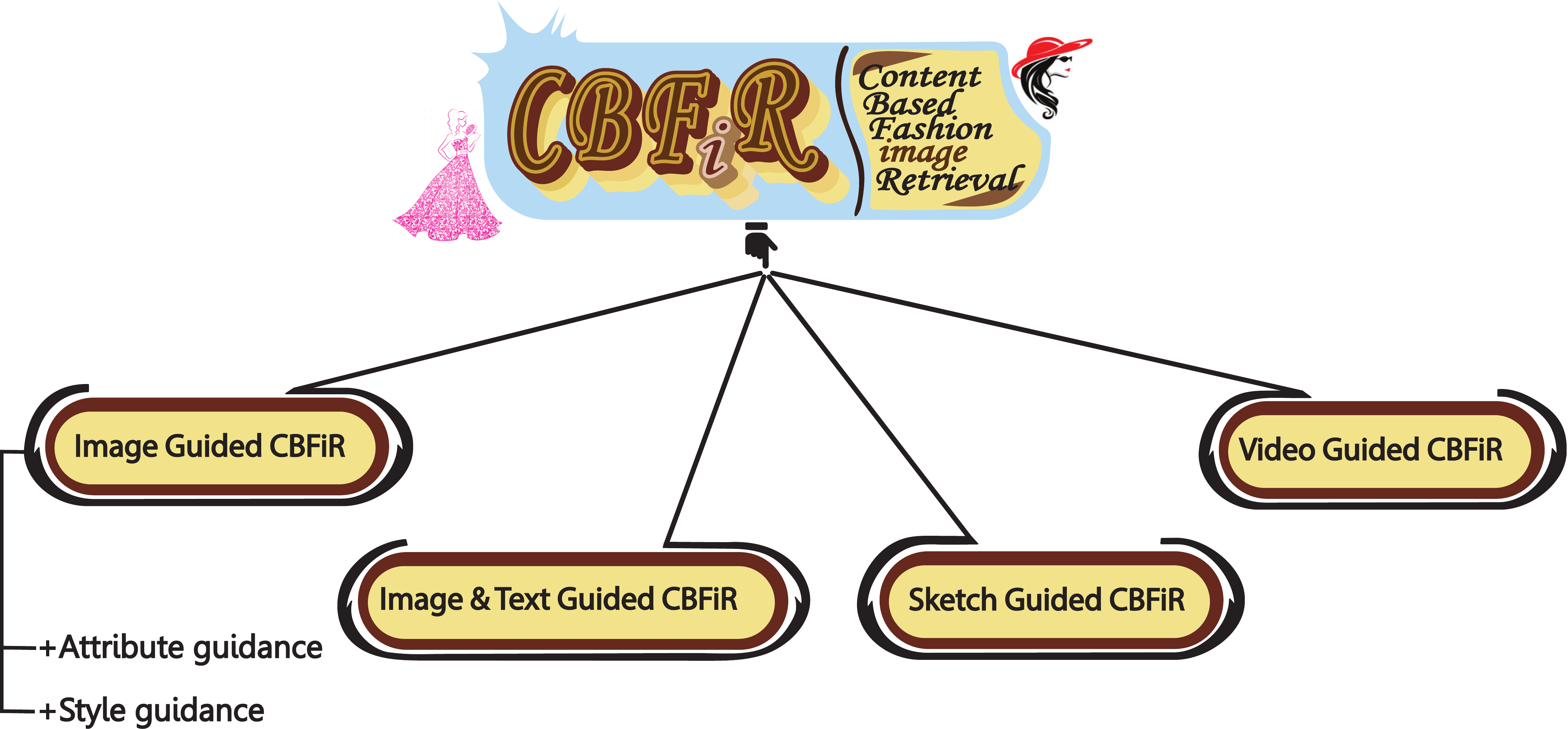}
 	\label{fig:taxonomyCBFIR}
 \end{figure}
 
 \section{Content-Based Fashion Image Retrieval (CBFIR) methods}
 \label{CBFIR-methods}
 
 Content-based fashion image Retrieval (CBFIR) methods retrieved the desired fashion items or products from the queried reference in the form of image, text, or visual clue. According to recent advancements in CBFIR, we divided the domain area into four major groups, i.e., image-guided CBFIR methods, image + text-guided CBFIR methods, sketch-guided CBFIR methods, and video-guided CBFIR methods. Figure \ref{fig:taxonomyCBFIR} shows the taxonomy diagram of these CBFIR methods.

 \subsection{Image-guided CBFIR}
\begin{figure}
 	\caption{The retrieved results of image-guided CBFIR methods according to the input queried image.}
 	\centering
 	\includegraphics[width=\textwidth]{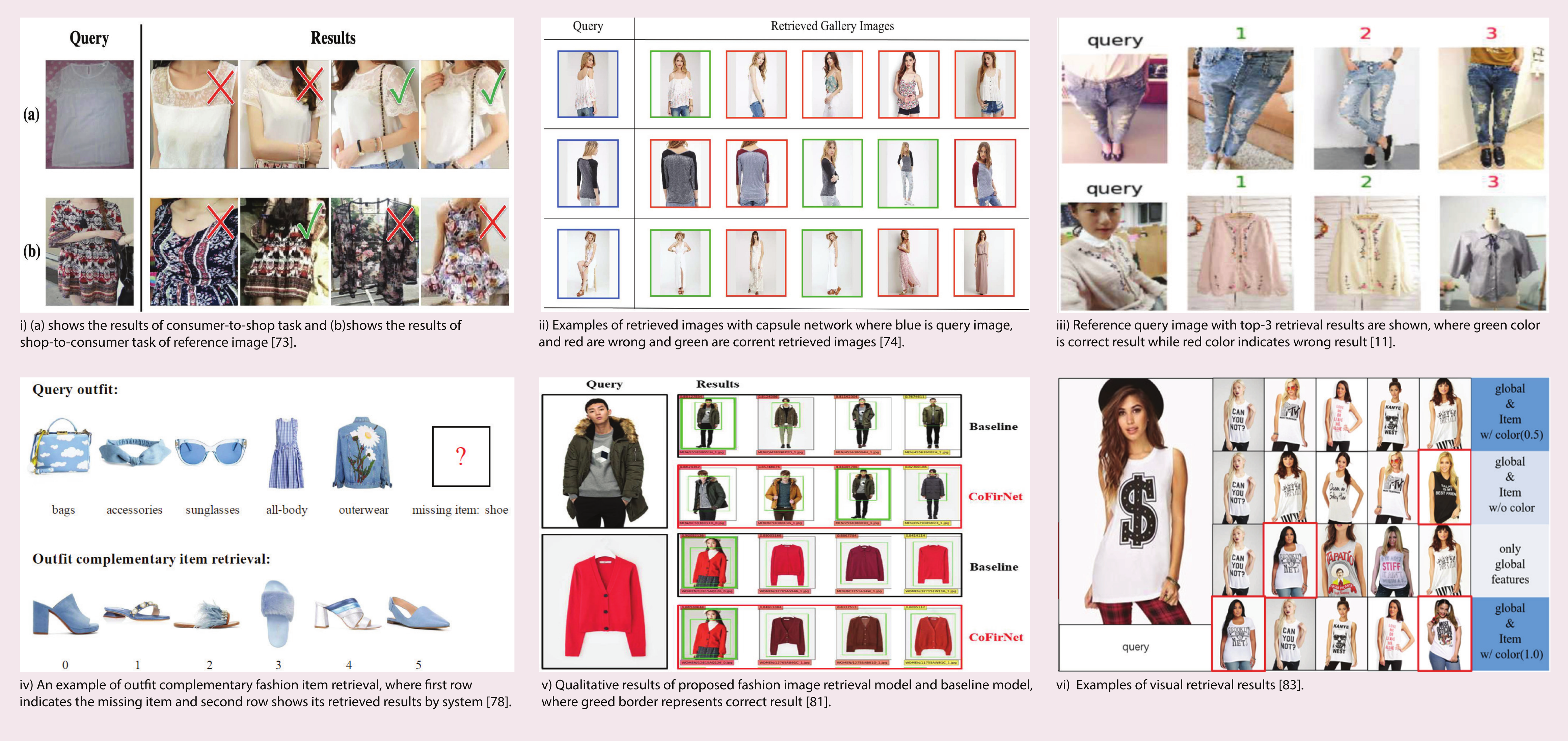}
 	\label{fig:image-G-CBFIR}
 \end{figure}

 In image-guided CBFIR methods, reference images are used to retrieve the desired fashion product for a user. There are two main processes in image-guided CBFIR, which have been investigated for years to find visually identical images. Similarity and feature vectors (FV) are two important mechanisms of image-guided CBFIR models. The FV is extracted to signify each reference and database image in the image-guided CBFIR system. The database and reference image similarity is calculated according to the extracted FVs. Therefore, the FVs and similarity of images play an important part in increasing the retrieval accuracy of image-guided CBFIR methods. During the real retrieval process, reference images supplied by users frequently experience issues, including bad lighting, posture variations, various shooting angles, and other considerations. The users' inability to supply perfect reference images as query input makes the content-based fashion image retrieval process more difficult. Figure \ref{fig:image-G-CBFIR} shows some results of retrieved images with different image-guided methods explained in this section with the perspective of the query reference image. In the recent development of image-guided CBFIR, the key contribution of authors are explained below and are analyzed comparatively according to CBFIR networks, publication year, datasets, evaluation metrics, and loss functions in Table \ref{tab:Image-guided-CBFIR}.
 
 \begin{table}[]
 	\scriptsize
 	\caption{Comparative analysis of Image-guided CBFIR methods, where mAP = mean Average Precision, N/A = Not Available}
 	\label{tab:Image-guided-CBFIR}
 	\begin{adjustbox}{width=1.1\textwidth, center=\textwidth}
 		\begin{tabular}{|p{2.6cm}|p{0.6cm}|p{3.2cm}|p{3.5cm}|p{2.3cm}|p{3.8cm}|}
 		\bottomrule
 		\rowcolor[HTML]{7CACF8}
 		\textbf{Model} & \textbf{Year} & \textbf{CBFIR Networks} & \textbf{Datasets} & \textbf{Evaluation Metrics} & \textbf{Loss Function} \\
 		\bottomrule
 		Ji, X. et al \cite{ji2017cross} & 2017 & CNN/SGD & DARN, DeepFashion & Top@K (5,20), precision & Triplet loss \\
 		Gajic, B. et al. \cite{gajic2018cross} & 2018 & CNN/ResNet50 & DARN, DeepFashion & Top@K (20) & Triplet loss \\
 		Kucer, M. et al. \cite{kucer2019detect} & 2019 & Mask R-CNN/FPN & ModaNet, Exact Street2Shop. & mAP, Top@K (1,20) & Triplet loss, average precision loss \\
 		Ibrahimi, S. et al. \cite{ibrahimi2019deep} & 2019 & CNN/ResNet50 & DeepFashion2 & Recall@K (1,5,10,20) & N-pair loss, lifted loss, angular loss, hard-triplet loss \\
 		Kinli, F. et al. \cite{kinli2019fashion} & 2019 & CNN & DeepFashion & Recall@K (1,10,20,30,40,50) & Triplet loss \\
 		GRNet \cite{kuang2019fashion} & 2019 & GCN/GoogleNet & DeepFashion,FindFashion, Street2Shop& Top@K (1,20,50) & Cross-entropy loss, classification loss \\
 		GSN \cite{chopra2019powering} & 2019 & CNN/Inception-v1 & DeepFashion, CARS196 & Recall@K (1,5,10,20,50) & Grid Search Loss \\
 		Park, S. et al. \cite{park2019study} & 2019 & CNN/DenseNet121 /ResNet50/SEResNet50 & DeepFashion & Top@K (5,20) & Cross-entropy loss, triplet loss \\
 		RST \cite{wieczorek2020strong} & 2020 & CNN/GoogLeNet /ResNet50 & DeepFashion, Street2Shop & mAP, recall@k (1,10,20,50) & Classification loss \\
 		CSA-Net \cite{lin2020fashion} & 2020 & CNN/ResNet18, CSA-Net, KNN & Polyvore Outfit & Recall@topk (10,30,50), AUC & Triplet loss, outfit ranking loss \\
 		Islam, SM. et al. \cite{islam2021ringfir} & 2020 & CNN & RingFIR & Recall@K (1,5,10) & N/A \\
 		PS-Net \cite{lang2020plagiarism} & 2020 & HR-Net, Faster R-CNN & DeepFashion2, Plagiarized Fashion & mAP, Recall@k (10,20,30) & Triplet ranking loss, plagiarized retrieval loss \\
 		CoFirNet \cite{ko2021cofirnet} & 2021 & CNN/PNASNet-5 & Manual dataset & Recall@K (5,10,15,20) & Triplet loss, multi-task loss, classification loss \\
 		Morelli, D. et al. \cite{morelli2021fashionsearch++} & 2021 & CNN/ResNet50-ResNet101 & Street2Shop & Recall@K (1,5,10,20) & Maximum hinge-based triplet ranking loss \\
 		FRSFC \cite{liu2021frsfn} & 2021 & SFN, CNN/Resnet50 & DeepFashion & Recall@k (3,5) & Softmax cross entropy, triplet loss \\
 		UMDA \cite{sharma2021unsupervised} & 2021 & CNN/ResNet50/ ResNet101 & DeepFashion, Street2Shop & mAP, Recall@K (1,5,10,20,50). & Triplet ranking loss\\
 		\bottomrule
 	\end{tabular}
	\end{adjustbox}
 \end{table}

 Ji, X. et al \cite{ji2017cross} proposed a fashion image retrieval framework using an attention modelling mechanism for the consumer and shop domain. Tag-based attention models are used to extract product features for fashion image retrieval tasks from both database images and the query image. A triplet ranking loss is used in this framework for metric learning of features. 
 Gajic, B. et al. \cite{gajic2018cross} created a three-stream Siamese framework for the fashion image retrieval task. The standard triplet loss function is used in this research to calculate the difference between query and database images. Weights of streams are initialized during the training phase using Siamese stream ranking. Learning rate and triplet mining at the training stage enhance the performance of the fashion image retrieval task. 
 Kucer, M. et al. \cite{kucer2019detect} introduced a detection and retrieval model for fashion images, where the query image is first processed through a detection model to detect a fine-grained fashion object. This detection model is specially designed to detect specific details about clothing items to minimize ambiguities. And then, triplet loss and AP loss boost the performance of the retrieval task. 
 Ibrahimi, S. et al. \cite{ibrahimi2019deep} proposed a deep learning architecture for fashion instance retrieval. In this research, the author labels every instance of fashion item according to its visibility perspective as occlusion, viewpoint, scale, and zoom-in. The model provides a baseline for shop-to-consumer and consumer-to-shop fashion instance retrieval tasks.

 Kinli, F. et al. \cite{kinli2019fashion} proposed architecture for the CBFIR task using capsule networks. Residual-connected (RC) and stacked-convolutional blocks are used to feed data to the capsule layers. The capsule networks are designed to recognize fashion items with fewer visual angles and without many transformations. This architecture learns higher dimensional configuration of poses from the input query image. 
 Kuang, Z. et al. \cite{kuang2019fashion} optimizes the performance of the CBFIR task using a graph reasoning network. This network learns the similarities between the database cloth image and query image using both the local and global representations. Different scales are used to represent the similarities between fashion items using a graph similarity pyramid. 
 Chopra, A. et al. \cite{chopra2019powering} proposed a grid search network using a feature embedding mechanism for fashion image retrieval tasks. The network's main emphasis is to find the exact matches of the query image in the system grid, which contains both the negative and positive cloth items according to the input image. The performance of the content-based retrieval task was improved by implementing a reinforcement learning technique on feature embeddings. 
 Park, S. et al. \cite{park2019study} examined the different training strategies and frameworks for CBFIR tasks to enhance their performance and accuracy. Instance-level labelling and category-level labelling are used for the training of CBFIR models. Instance level labelling helps to label each cloth item detected by the model. 
 Wieczorek, M. et al. \cite{wieczorek2020strong} redesign a person Re-Identification model to deal with the problems of background clutters, quality gaps, different lighting conditions, and contrast in CBFIR tasks from consumer and shop database domains. Using different training skills, the author achieves better performance and accuracy for fashion item retrieval tasks on Street2Shop and DeepFashion datasets. 
 
 Lin, YL. et al. \cite{lin2020fashion} proposed a model that retrieves compatible fashion items according to available fashion items available. Using this research scheme, a consumer will easily purchase a new desired target fashion item by matching it with existing items available. The CNN network is trained to generate feature embeddings for every fashion item of an outfit using query image, its category, and target category. An output ranking loss enhances the outfit retrieval accuracy as well as a compatible prediction for CBFIR. 
 Islam, SM. et al. \cite{islam2021ringfir} proposed a dataset for earrings image retrieval in the fashion domain. Different baseline retrieval methods are tested on the proposed dataset to evaluate its performance. Compared to other CBFIR methods, the earring image retrieval method is a more challenging task because of too much variation in its structure variety, patterns, and designs. 
 Lang, Y. et al. \cite{lang2020plagiarism} proposed a framework that uses specified regions of plagiarized clothes to enhance the performance of fashion image retrieval tasks. For this specific task, the author designed a new dataset for clothes retrieval named "Plagiarized Fashion." A regional attention mechanism is used to extract the region-based features for improved content-based fashion image retrieval. 
 Ko, MS. et al. \cite{ko2021cofirnet} proposed a vector-based FIR architecture, where feature vectors are extracted from query images using domain-specific knowledge. The conditional triplet and classification networks enhance the FIR task performance by combining the specific feature vectors extracted using the feature extractor block. 
 
 Morelli, D. et al. \cite{morelli2021fashionsearch++} proposed a content-based fashion image retrieval framework using a modified triplet loss mechanism at the training phase. The performance of cloth image retrieval is improved by considering the hard negative samples at the training. The experimental results show improved results on fashion categories and fashion accessories compared to baseline methods. 
 Liu, AA. et al. \cite{liu2021frsfn} proposed a semantic fusion-based network for fashion image retrieval tasks. This architecture semantically extracts the item and global features from the input reference image. Semantic fusion using color information on these extracted features enhances the accuracy of cloth image retrieval. 
 Sharma, V. et al. \cite{sharma2021unsupervised} proposed an unsupervised adaptation method for fashion image retrieval. Training is difficult for all those cloth images whose respective consumer images are unavailable in the target database. To overcome the meta-domain gap for this scenario, the framework used a similar type of images and did some pseudo-labelling for training missing consumer images.

 \subsubsection{Attribute guidance}
 
\begin{figure*}
 	\caption{The retrieved results of image-guided CBFIR with attribute guidance methods according to the input queried image and desired attribute in textual form.}
 	\centering
 	\includegraphics[width=\textwidth]{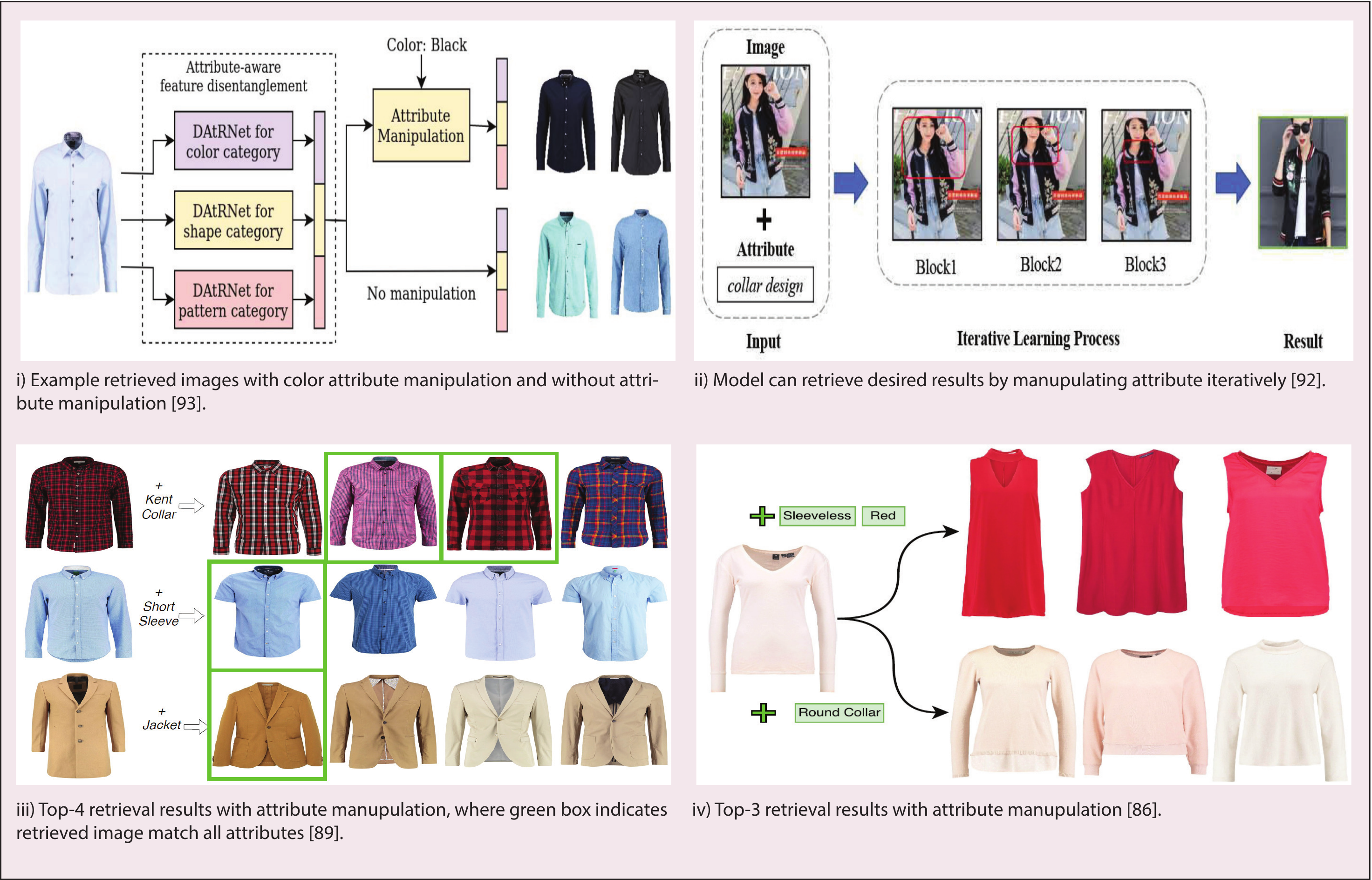}
 	\label{fig:image-G-Attribute}
 \end{figure*}

 Attribute guidance plays an important role in retrieving some specific cloth images with the desired attributes by the consumers, like a particular color, style, pattern, and many more. A main concern to be addressed is that fashion products or items have many attributes, and it is vital for each attribute to have representative features. Attribute manipulation is also challenging in content-based fashion image retrieval because combining query image fashion features and desired attribute representations is a very complex process. Attribute guidance positively impacts fashion retrieval tasks and leaves too much room for future researchers. Figure \ref{fig:image-G-Attribute} shows some results of retrieved images with different image-guided CBFIR methods with attribute guidance with the perspective of the query reference image and desired attributes in text form. In the recent development of image-guided CBFIR with attribute guidance, the key contribution of authors are explained below and are analyzed comparatively according to CBFIR networks, publication year, datasets, evaluation metrics, and loss functions in Table \ref{tab:Image-attribute-style CBFIR}.
 
 \begin{table}[]
 	\caption{Comparative analysis of Image-guided CBFIR methods with attribute and style guidance, where BCE = binary cross entropy, mAP = mean Average Precision, CL = Classification Loss, TL = Triplet Loss, CEL = Cross-Entropy Loss, AL = Adversarial Loss}
 	\label{tab:Image-attribute-style CBFIR}
 	\scriptsize
 	\begin{adjustbox}{width=1.1\textwidth, center=\textwidth}
 	\begin{tabular}{|p{1.3cm}|p{2cm}|p{0.5cm}|p{2.6cm}|p{3.5cm}|p{2.2cm}|p{3.3cm}|}
 		\bottomrule
 		\rowcolor[HTML]{7CACF8}
 		& \textbf{Model} & \textbf{Year} & \textbf{CBFIR Networks} & \textbf{Datasets} & \textbf{Evaluation Metrics} & \textbf{Loss Function} \\
 		\bottomrule
 		\multirow{8}{*}{\cellcolor[HTML]{adcaf8} } & FashionSearchNet \cite{ak2018learning} & 2018 & CNN/AlexNet & Shopping100k, DeepFashion & Top-K (30) & CL, triplet ranking loss, global ranking loss \\
 		\cellcolor[HTML]{adcaf8}& FLAM \cite{shin2019semi} & 2019 & GAN & DARN, Shopping100k & Recall@K (1,5,20,50) & TL, AL, feature matching loss. Cycle consistency loss \\
 		\cellcolor[HTML]{adcaf8}& FashionSearchNet-v2 \cite{ak2021fashionsearchnet} & 2021 & CNN/AlexNet & Shopping100k, DARN, iMaterialist, CelebA & Recall@K (20) & CL, triplet ranking loss, global ranking loss \\
 		\cellcolor[HTML]{adcaf8} +Attribute & AHBN \cite{su2020look} & 2021 & Inception-ResNet-v2, hourglass network & Exact Street2Shop, DeepFashion & Recall@K (1, 10, 20, 30, 40, 50) & BCE loss, landmark loss, CEL \\
 		\cellcolor[HTML]{adcaf8}& Hou, Y. et al. \cite{hou2021learning} & 2021 & CNN/AlexNet/ResNet & Shopping100k, DeepFashion & Recall@K (10, 20, 30, 40, 50) & CEL, compositional TL, consistency loss, Label TL. \\
 		\cellcolor[HTML]{adcaf8}& D'Innocente, A. et al. \cite{d2021localized} & 2021 & CNN/ResNet50 & FashionLocalTriplets & Precision, Recall@K(1,5,10) & Localized triplet loss \\
 		\cellcolor[HTML]{adcaf8}& ISLN \cite{yan2022attribute} & 2022 & CNN/ResNet50 & FashionAI, DARN & mAP & TL \\
 		\cellcolor[HTML]{adcaf8}& DAtRNet \cite{bhattacharya2022datrnet} & 2022 & CNN & DeepFashion, Shopping100k & Recall@K (30) & TL \\
 		\bottomrule
 		\multirow{2}{*}{} \cellcolor[HTML]{d5e3f8} & Comp-Net \cite{valle2018effective} & 2018 & CNN & Chictopia web data & Top@K (1,5,10) & cost-sensitive loss \\
 		\cellcolor[HTML]{d5e3f8} +Style & Naka, R. et al. \cite{naka2022fashion} & 2022 & CNN/ResNet50, U2-Net & FashionStyle14, iMat-Fashion & mAP & Quadruplet loss\\
 		\bottomrule
 	\end{tabular}
 \end{adjustbox}
 \end{table}

 AK, KE. et al. \cite{ak2018learning,ak2018efficient} proposed a model that takes attributes and query images for retrieving similar cloth images. The model extracts the query image attributes and manipulates some specific attributes like collar and color attributes. Attribute activation maps are explicitly used to represent and extract the most related region of cloth image. Therefore, the main research perspective of focusing on specific regions of cloth images increases the performance of fashion image retrieval using these attribute representations. 
 Shin, M. et al. \cite{shin2019semi} proposed an attribute manipulation framework using a feature-level semi-supervised learning technique for the CBFIR task. This research uses a Generative Adversarial Network (GAN) architecture to manipulate the color, pattern, and shape attributes of a fashion item. This framework manipulates the learned feature attributes directly at the feature level with less utilization of computational power. The model training at the manipulation and embedding stages increases the performance of the fashion retrieval task with feature-level attribute guidance. 
 Su, H. et al. \cite{su2020look} proposed an attention mechanism-based bilinear network for the CBFIR task. This heterogeneous network uses one branch to take appearance attributes from the query image, and the other branch is used to take localization information using the landmark detection mechanism. These extracted heterogeneous features are again finetuned using the joint attentional mechanism to improve the accuracy of fashion image retrieval tasks. 
 Ak, KE. et al. \cite{ak2021fashionsearchnet} designed a FashionSearchNet-v2 framework for specific attribute manipulation. This framework extracts specific representations from attributes using a weekly supervised localization block to improve similarity learning. A combination of triplet ranking and classification loss estimates the local representations at the training stage. All the local representations extracted at the training stage are combined in a common space to improve the fashion item retrieval prediction with desired attribute manipulation.
 
 Hou, Y. et al. \cite{hou2021learning} proposed a model that retrieves interactive fashion items according to specific attributes provided by consumer feedback. This model can change the specific attribute type by preserving the actual fashion cloth representation, like changing the color of the fashion item, changing the sleeve type, changing in pattern, retrieving different cloth items with similar colors, etc. 
 D'Innocente, A. et al. \cite{d2021localized} proposed a framework to deal with the aggregated similarities issues at the local level representations for fashion image retrieval. This framework can learn the local interest features of user-specified prominent visual features. The localized triplet loss is modified accordingly using desired attributes to enhance the accuracy of cloth retrieval. A specified FashionLocalTriplets dataset of 4,302 female dress images is also proposed to test the framework's accuracy. 
 Yan, C. et al. \cite{yan2022attribute} proposed a content-based fashion image retrieval model using attribute guidance by an iterative similarity learning mechanism. In this model, a query image is put into the system along with some attribute values to leverage the more accurate features by focusing on the specific area of interest. This model can improve feature learning using both the part-level and whole-level features during different iterations. 
 Bhattacharya, G. et al. \cite{bhattacharya2022datrnet} designed a DAtRNet network for content-based fashion image retrieval. DAtRNet disentangles query image features according to color, pattern, and shape attributes. Attributes are manipulated with the perspective of these extracted attributes from color, pattern, and shape blocks. The axial attention scheme re-calibrates the soft attention masks adaptively. The network is able to deal with or separate the attribute features independently from one or multiple attributes. 
 
 \subsubsection{Style guidance }
 
 Style guidance plays a vital role in content-based fashion image retrieval tasks, as a retrieved fashion image for the wedding may not be suitable for the graduation ceremony. Style, occasion, and season are some key examples of the cultural and environmental factors that make up an outfit's semantics. The retrieved image is considered as a perfect match if the reference and retrieved images are linked with the same style and suitable for the same seasons and events. In the recent development of image-guided CBFIR with style guidance, the key contribution of authors are explained below and are analyzed comparatively according to CBFIR networks, publication year, datasets, evaluation metrics, and loss functions in Table \ref{tab:Image-attribute-style CBFIR}.
 
 Valle, D. et al. \cite{valle2018effective} proposed a CBFIR using semantic compositional networks considering the occasion, season, and style concepts. To minimize the semantic gap between the query and retrieved fashion images, a conditional normalization layer is used along with the SoftMax function. A SkipGram representation is used to learn the features associated with occasion, season, and style tags/labels.  
 Naka, R. et al. \cite{naka2022fashion} proposed a style-aware fashion image retrieval framework for CBFIR. Fashion images are retrieved using style descriptions extracted using user posts like style tags, season tags, consumer height, silhouette, etc. The framework learns embeddings using the quadruplet loss function that deliberates the style-aware description features and similarity-based pairing of visual features. These multiple-choice-based style-aware consumer descriptions retrieved the fashion images according to specific needs.

 \subsection{Image \& text-guided CBFIR}
 
\begin{figure*}
 	\caption{The retrieved results of image \& text guided CBFIR methods according to the input reference image and desired attribute in textual form.}
 	\centering
 	\includegraphics[width=\textwidth]{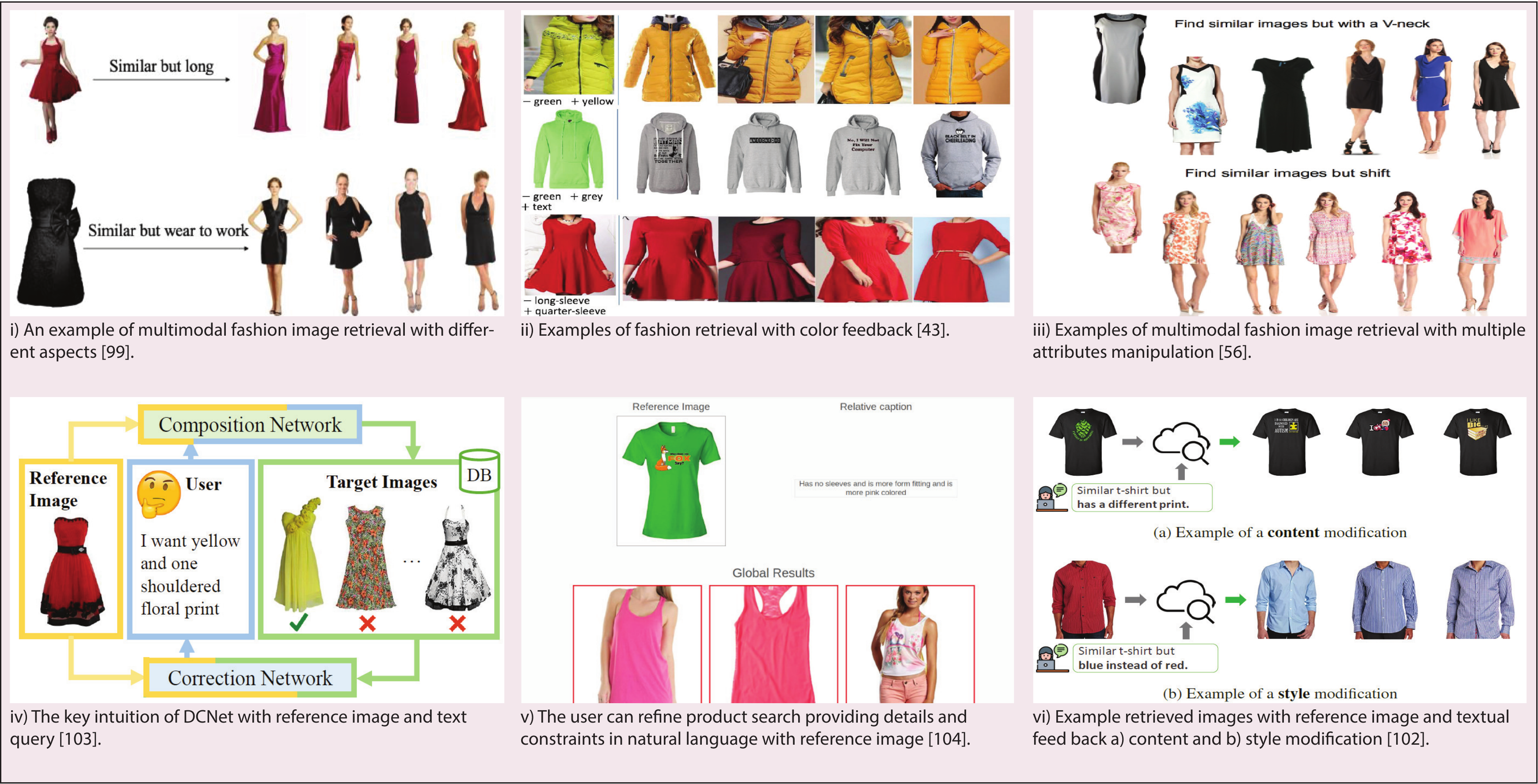}
 	\label{fig:image&text CBFIR}
 \end{figure*}

 In traditional searching schemes for fashion images, users put keywords in search engines to find fashion items of their desire. Users can retrieve only those products whose keywords match the meta-data of fashion items, like fashion item titles, tags, descriptions, etc. Therefore, it is a very challenging task to label such a huge amount of fashion item data. To deal with this scenario, many researchers proposed image and text-guided CBFIR models to enhance the accuracy of clothing item and product retrieval tasks. Feature representations of query images and textual keywords together represent some dominant results in CBFIR tasks. In recent years, with the advancement of deep learning procedures, the authors proposed various methods using image \& text guidance for CBFIR. Figure \ref{fig:image&text CBFIR} shows some results of retrieved images with different image \& text-guided CBFIR methods with the perspective of the query reference image and desired characteristics about fashion items in text form. In the recent development of image \& text-guided CBFIR, the key contribution of authors are explained below and are analyzed comparatively according to CBFIR networks, publication year, datasets, evaluation metrics, and loss functions in Table \ref{tab:Image and  text-guided CBFIR}.
 
 \begin{table}[]
 	\caption{Comparative analysis of Image \& text-guided CBFIR methods, where mAP = mean Average Precision, CL = Contrastive Loss, TL = Triplet Loss, CEL = Cross-Entropy Loss, RL = Ranking Loss, N/A = Not Available}
 	\scriptsize
 	\label{tab:Image and  text-guided CBFIR}
 	\begin{adjustbox}{width=1.1\textwidth, center=\textwidth}
 	\begin{tabular}{|p{2cm}|p{0.5cm}|p{3.3cm}|p{3.5cm}|p{2.6cm}|p{3cm}|}
 		\bottomrule
 		\rowcolor[HTML]{7CACF8}
 		\textbf{Model} & \textbf{Year} & \textbf{CBFIR Networks} & \textbf{Datasets} & \textbf{Evaluation Metrics} & \textbf{Loss Function} \\
 		\bottomrule
 		Corbiere, C. et al. \cite{corbiere2017leveraging} & 2017 & CNN/ResNet50 & DeepFashion,Manual dataset & Top@K (3,5) & CEL \\
 		Rubio, A. et al. \cite{rubio2017multi} & 2017 & CNN/AlexNet, word2vec & e-commerce & Recall@k (5,10) & CL \\
 		Liao, L. et al. \cite{liao2018interpretable} & 2018 & CNN/ResNet50, RNN/BLSTM, GAP & DARN, Amazon & Top@K (1), Recall@K (10) & Bi-directional RL, CEL \\
 		Laenen, K. et al. \cite{laenen2018web} & 2018 & CNN/CaffeNet & Amazon dataset & mAP, precision@K (1,5) & N/A \\
 		Gu, X. et al. \cite{gu2018multi} & 2019 & CNN/VGG-16, SGD & Manual dataset & mAP, NDCG & Quintuplet-based RL, cross-view similarity RL \\
 		FashionBERT \cite{gao2020fashionbert} & 2020 & CNN/InceptionV3 /ResNeXt101, BERT & Fashion-Gen & Rank@K (K=1, 5, 10) & Adaptive loss, CEL \\
 		FashionCorrNet \cite{yan2020joint} & 2020 & CNN/ResNet152, Word2vec & Amazon dataset & Euclidean distance, Manhattan distance, Chebyshev distance & mean square error loss \\
 		Baldrati, A. et al. \cite{baldrati2021conditioned} & 2021 & CLIP/ResNet-50/ ResNet-50x4 & FashionIQ & Recall@K (5,10) & CL \\
 		Yuan, Y. et al. \cite{yuan2021conversational} & 2021 & GloVe, GRU, CNN/ ResNet-101/ResNet-152 & FashionIQ & Recall@K (5,8), MRR & Max-margin triple loss \\
 		CoSMo lee2021cosmo & 2021 & CNN/ResNet18/ ResNet50, RNN/LSTM & FashionIQ, Fashion200K & Recall@K (1,10,50) & Batch-based classification loss \\
 		DCNet \cite{kim2021dual} & 2021 & CNN/ResNet50, GloVe & FashionIQ, Fashion200K & Recall@K (10,50) & CEL \\
 		Baldrati, A. et al. \cite{baldrati2022effective} & 2022 & CNN/ResNet50, CLIP text encoder & FashionIQ, CIRR & Recall@K (1,5,10,50) & Batch-based classification loss \\
 		FashionVLP \cite{goenka2022fashionvlp} & 2022 & CNN/ResNet18/ ResNet50, BERT & FashionIQ, Fashion200K & Recall@K (10,50) & Batch-based classification loss \\
 		\bottomrule
 	\end{tabular}
 \end{adjustbox}
 \end{table}

 Corbiere, C. et al. \cite{corbiere2017leveraging} proposed a CBFIR model on the bases of a weekly supervised learning scheme using query image and text guidance as input. This model learns representations from data available on e-commerce websites catalogs; that's why this model doesn't need manual labelling of data. The model is capable of dealing with multi-languages, like, English, French, and Italian. This model achieves optimum results for attribute prediction and fashion image retrieval. 
 Rubio, A. et al. \cite{rubio2017multi} designed a model that jointly embeds multimodalities (i.e., image and text) for fashion image retrieval. A common latent space is used to learn the representations from both text and image queries. The joint embedding accuracy is increased by training with specified classification networks and large daily-life e-commerce fashion product data available. The similarity distance between related text and image outcome is minimized, while unrelated text and image outcome is maximized using this scheme. The contrastive loss function is used with specified classification networks to retain accurate semantic information. 
 Liao, L. et al. \cite{liao2018interpretable} designed a novel Excusive and Independent (EI) tree structure that collaborates with deep learning models for image and text modalities learning. The proposed tree structure categorizes the fashion concepts into different hierarchies semantically and expends the EI tree with independent and exclusive constraints. The model can learn in an end-to-end manner with these sub-categories of the tree structure. With the EI tree hierarchies, the model can retrieve with respect to fashion items concepts feedback and automatic retrieval as well.   
 Ak, KE. et al. \cite{ak2018learning} proposed a model that learns local representations of fashion items using specific regions. The proposed FashionSearchNet framework uses a weakly supervised learning scheme to extract specific areas of fashion items using queried attributes. Similarity learning can be improved using only specific regions of fashion items. Attribute activation maps are explicitly used to represent and extract the most related region of cloth image.
 
 Laenen, K. et al. \cite{laenen2018web} proposed a model for the CBFIR task that uses query image and text as input to retrieve the desired fashion product with slight variation in the perspective of the queried text. The model ranks the similarity of retrieved fashion images based on intermodal (image, text) representations in a common multimodal space. The model can easily manipulate the attributes of the query image with the perspective of the queried text. 
 Gu, X. et al. \cite{gu2018multi} proposed a CBFIR framework that not only learns representations from different modalities (image and text) but also learns representations from different domains. This framework works well for fashion image retrieval and analysis using heterogeneous fashion data. The accuracy of fashion image retrieval is enhanced using this model because it jointly considers heterogeneous and homogeneous similarity constraints.
 Gao, D. et al. \cite{gao2020fashionbert} proposed the FashionBERT framework for cross-modal fashion image retrieval using image and text modalities. Generally, text descriptions (like color, style, pattern, attributes, etc.) about fashion products provide more information than the region of interest-based product information. This framework used patches from the query image and product attributes from its description to accurately learns the features for the retrieval task. The adaptive loss function plays an important role in enhancing fashion image retrieval accuracy. 
 Yan, C. et al. \cite{yan2020joint} designed a FashionCorrNet framework for CBFIR using image and text modalities. The FashionCorrNet framework uses a correlational neural network to optimize the semantic similarity among various fashion representations of queried image and text modalities. The text part as input modifies the semantic features of retrieved images according to consumer desire. Additionally, this framework can easily achieve the coss-modal conversion. 
 
 Baldrati, A. et al. \cite{baldrati2021conditioned} proposed a framework that used a Contrastive Language-Image Pre-training (CLIP) model for conditional fashion image retrieval using the contrastive learning scheme. A combiner CLIP network is used to learn the representations of the query image and additional textual description of what the consumer wants. These combiner networks increase the accuracy of conditional retrieval of fashion images. Additionally, image and text scalar weights control the similarity ranking between modalities. 
 Yuan, Y. et al. \cite{yuan2021conversational} designed a framework for CBFIR, which uses multiturn textual feedback from consumers to retrieve the desired fashion image. Most baseline methods use single feedback for manipulating fashion image attributes, but multiturn textual feedback re-retrieves the target image based on consumer turn-by-turn textual feedback. Additionally, the attribute information of fashion images is leveraged through a mutual attention mechanism. 
 Lee, S. et al. \cite{lee2021cosmo} proposed a Content-Style Modulation (CoSMo) framework for CBFIR using the image and text modalities. In this research, the framework can manipulate the content and style of retrieved fashion images based on the text query. The CoSMo designed two different modules, i.e., content modulator and style modulator, for manipulating content and styles from queried images and text. Desired content modifications are performed by the content modulator on the reference image and update the features accordingly. Then global style representations are updated by the style modulator.
 Kim, J. et al. \cite{kim2021dual} proposed the Dual Composition Network (DCNet) for retrieving target fashion images based on the reference image and textual query. The composition and correction networks learn the compositional representations from queried images and text to retrieve the desired target image. Both these composition networks and correction network works in a close loop to retrieve the desired image more accurately. 
 Baldrati, A. et al. \cite{baldrati2022effective} proposed a CBFIR model using conditional query and reference image, which is very useful for e-commerce applications. Additional requirements and intent of the user are fed into the system through textual feedback to finetune the retrieved fashion image search. This research's key application is to enhance e-commerce search engines' performance in the fashion image retrieval domain. Users can easily change the attributes of searched fashion items by providing textual feedback, like changing the color of the product, changing the pattern, changing the logo, etc. 
 Goenka, S. et al. \cite{goenka2022fashionvlp} designed a FashionVLP framework for fashion product retrieval using text and image modalities. Similarly, this framework also takes feedback from the user in textual form to finetune the target retrieved image. The model's accuracy is improved by designing the designated attention mechanism for fusing target fashion product features without concerning the transformer layers. The FashionVLP leverage the existing knowledge from the huge image-text database for the retrieval task using consumer feedback in textual form.
 
 \subsection{Sketch-guided CBFIR}
 
\begin{figure*}
 	\caption{The retrieved results of sketch-guided CBFIR methods according to the input sketch as a reference image.}
 	\centering
 	\includegraphics[width=\textwidth]{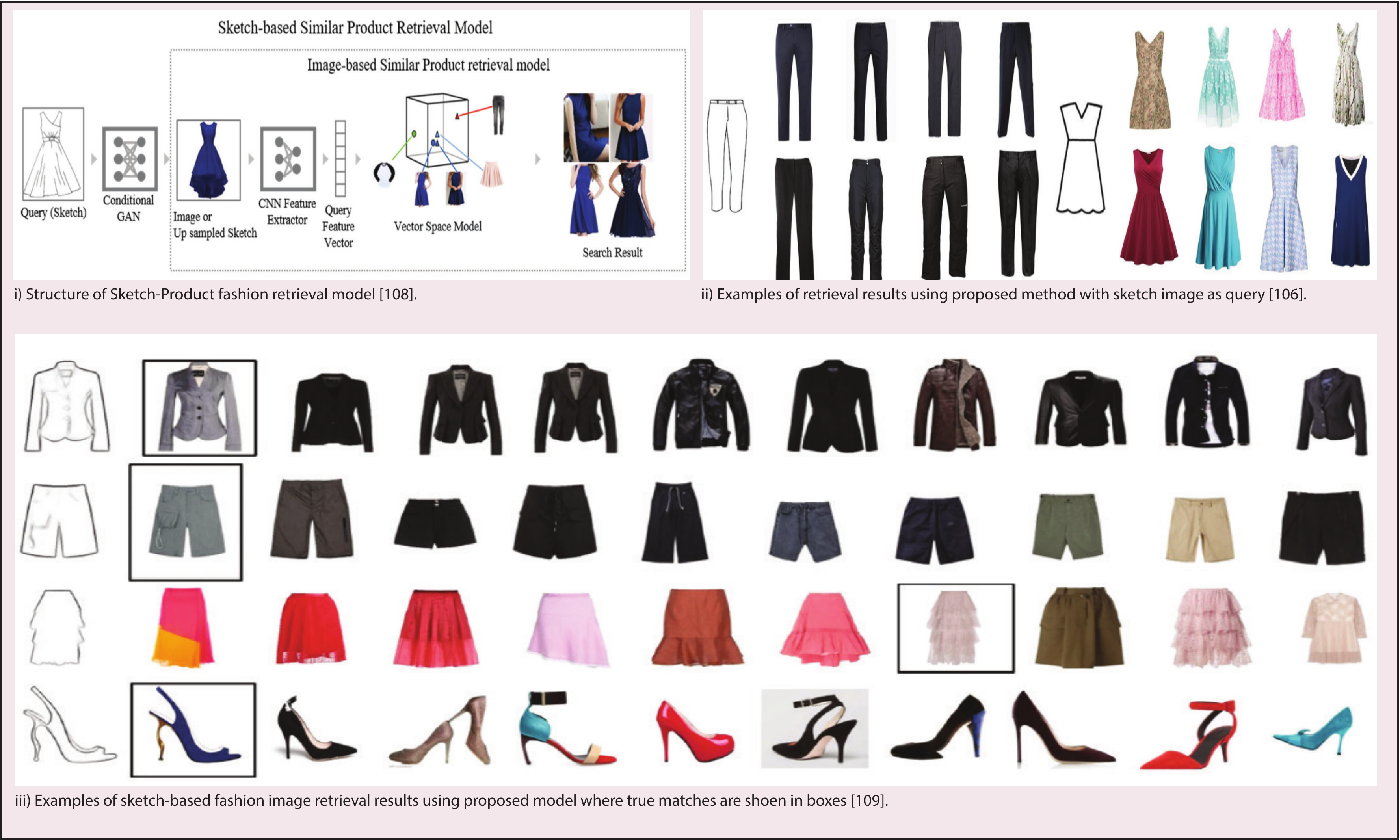}
 	\label{fig:sketch CBFIR}
 \end{figure*}

 In sketch-guided CBFIR methods, reference sketch images are used to retrieve a user's desired fashion item or product. Sketch-guided CBFIR methods, in contrast to image-guided and image \& text-guided fashion image retrieval method, offers users a more intuitive and natural means of expressing their search requirements. A user can retrieve desired fashion products with specific requirements using sketch-based queries. Finding fashion items or products using free-hand sketches is not an easy task because of the considerable cross-domain disparity between sketches and fashion item images. Compared to fashion product images, fashion product sketches are more abstract and lack details like clothing colors, materials, and patterns, which makes it more challenging to retrieve garments with finer detail. Nevertheless, fashion image retrieval with sketch guidance has some key advantages, i.e., 1) sketches for fashion items or product cover more content as compared to text, 2) fashion product sketches can easily express the style of the desired product without any ambiguity, and 3) fashion image retrieval with sketch guidance is easy to obtain as compared to image-guided CBFIR. Fashion items or product retrieval with sketch guidance is still a growing field. Researchers are consequently significantly motivated to propose more efficient sketch-guided CBFIR approaches that use sketch images as reference queries to retrieve desired fashion products. Figure \ref{fig:sketch CBFIR} shows some results of retrieved images with different sketch-guided CBFIR methods explained in this section with the perspective of the query sketch image. In the recent development of sketch-guided CBFIR, the key contribution of authors are explained below and are analyzed comparatively according to CBFIR networks, publication year, datasets, evaluation metrics, and loss functions in Table \ref{tab:sketch-guided CBFIR}.
 
 \begin{table*}[]
 	\caption{Comparative analysis of Sketch-guided CBFIR methods, where mAP = mean Average Precision, CL = Consistency Loss, CEL = Cross-Entropy Loss, N/A = Not Available}
 	\label{tab:sketch-guided CBFIR}
 	\begin{adjustbox}{width=1.1\textwidth, center=\textwidth}
 	\begin{tabular}{|p{2.4cm}|p{0.6cm}|p{3.3cm}|p{3.5cm}|p{2.3cm}|p{3cm}|}
 		\bottomrule
 		\rowcolor[HTML]{7CACF8}
 		\textbf{Model} & \textbf{Year} & \textbf{CBFIR Networks} & \textbf{Datasets} & \textbf{Evaluation Metrics} & \textbf{Loss Function} \\
 		\bottomrule
 		Lei, H. et al. \cite{lei2018new} & 2018 & CRF, Bag-of-Words & TU Berlin, manual dataset & mAP & N/A \\
 		Chen, S. et al. \cite{chen2020cross} & 2020 & GAN, VAE, CNN/VGG16 & Manual dataset & Top@K (1,10) & CL \\
 		Jo, J. et al. \cite{jo2020development} & 2020 & GAN, CNN & Manual dataset & Precision & Point-wise ranking loss \\
 		Lei, H. et al. \cite{lei2021new} & 2021 & VAE, GAN, CNN/VGG16 & Fashion Image, QMUL-shoes, QMUL-chairs & Recall@K (1,10) & Reconstruction loss, CEL\\
 		\bottomrule
 	\end{tabular}
\end{adjustbox}
 \end{table*}

 Lei, H. et al. \cite{lei2018new} proposed a novel framework for fashion image retrieval using sketch reference images. Mobile visual sensors play an important role in drawing sketches as input for the retrieval of required fashion products. A conditional random field module is designed to classify the segments of query sketch images into multiple fragments. For training of the model, many fashion product images and their sketches are tagged manually with semantic component labels. Many future descriptors are designed to distinguish and capture each component. Dynamic component weighting scheme improves the accuracy of retrieving tasks using local and global features from the sketch-image query. 
 Chen, S. et al. \cite{chen2020cross} proposed a CBFIR framework for fine-grained fashion product retrieval using sketch guidance. The author designed a new sketch-photo pair dataset for the proposed framework using more than 34 thousand combinations under 26 different fashion image categories. This model uses VGG16 and unsupervised image-to-image translation networks (UNIT) to retrieve the target fashion image. The UNIT architecture deals with the heterogeneous nature of clothing and sketches images, and also rich annotated dataset is not required. 
 Jo, J. et al. \cite{jo2020development} proposed a sketch-based fashion image retrieval model that takes a user sketch as input for the desired fashion item, i.e., up-sampled using a GAN architecture. After that, the properties of that up-sampled fashion item are extracted as vector values and are used to find a similar item in the vector space. The cosine similarity measurement technique retrieves the similar fashion item based on these extracted vector values. 
 Lei, H. et al. \cite{lei2021new} proposed a novel cross-domain fashion image retrieval model using sketch-image guidance. The proposed model transforms images and sketches into the same domain. Image-based and sketch-based similarity are considered the key components to enhancing the retrieval accuracy of target fashion products. More than 36 thousand image-sketch pairs are labeled to calculate accuracy of the proposed model. Intensive experiments show the improved results on the proposed dataset along with QMUL-chairs and QMUL-shoes datasets.

 \subsection{Video-guided CBFIR}
 
\begin{figure*}
 	\caption{The retrieved results of video-guided CBFIR methods according to the reference short clips/videos.}
 	\centering
 	\includegraphics[width=\textwidth]{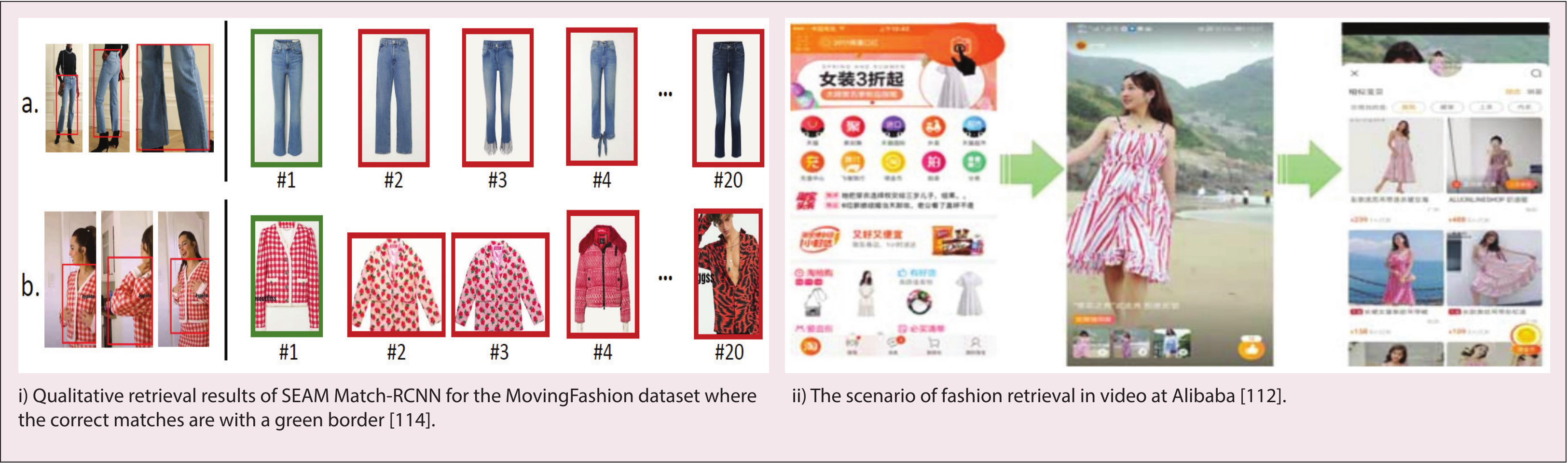}
 	\label{fig:video CBFIR}
 \end{figure*}

 Content-based fashion image retrieval using video guidance is another important category, where the researchers extract feature vectors from visual streams instead of images, text, or sketches, and similarity ranking has been calculated from the input visual stream for the retrieval of desired fashion products. Although television shows and movies are ideal venues for clothing companies to advertise their fashion items, viewers are not always fully conscious of where to purchase the recent fashion trends they view on screen. These television shows and movies can range millions of personalities worldwide and display the latest fashion trends. Additionally, social visual platforms like YouTube, Facebook, TikTok, Youku, Instagram, etc., have numerous users posting, viewing, and sharing billion of fashion-related videos every day. Clothing companies want to promote their fashion products using these visual platforms all over the world. However, buying fashion products from the video is a challenging task. Whenever a consumer wants to buy a trendy shoe pair or fancy dress in a video or film, there is not enough information available to accomplish the purchase. Therefore, searching for the desired product from a video or film is challenging. Recently, researchers put their efforts into dealing with this scenario in the domain of fashion image retrieval from visual steams. Figure \ref{fig:video CBFIR} shows some results of video-guided CBFIR methods with a short clip/video guidance. In the recent development of video-guided CBFIR, the key contribution of authors are explained below and are analyzed comparatively according to CBFIR networks, publication year, datasets, evaluation metrics, and loss functions in Table 5.
 
 \begin{table*}[]
 	\caption{Comparative analysis of Video-guided CBFIR methods, where mAP = mean Average Precision, TL = Triplet Loss, CEL = Cross-Entropy Loss, N/A = Not Available}
 	\label{tab:video-guided CBFIR}
 	\begin{adjustbox}{width=1.1\textwidth, center=\textwidth}
 	\begin{tabular}{|p{3cm}|p{0.6cm}|p{2.3cm}|p{3.5cm}|p{2.3cm}|p{3cm}|}
 		\bottomrule
 		\rowcolor[HTML]{7CACF8}
 		\textbf{Model} & \textbf{Year} & \textbf{CBFIR Networks} & \textbf{Datasets} & \textbf{Evaluation Metrics} & \textbf{Loss Function} \\
 		\bottomrule
 		Garcia, N. et al. \cite{garcia2017dress} & 2017 & CNN & Manual dataset & Accuracy & N/A \\
 		DPRNet \cite{zhao2021dress} & 2021 & CNN/RPN & DeepFashion, DeepFashion2, Video-to-Shop & Recall@K (1, 20, 50) & TL, classification loss, similarity loss \\
 		VACANet \cite{zhang2021exploring} & 2021 & CNN/ResNet50, BERT & Tmall Fashion Retrieval & Recall@k (1,20) & TL, quality domain alignment loss \\
 		Fashion Focus \cite{zhang2021fashion} & 2021 & ASR, CNN & Manual dataset & Recall & N/A \\
 		Godi, M. et al. \cite{godi2022movingfashion} & 2022 & Match-RCNN & MovingFashion & Top@K (1,5,10,20) & CEL\\
 		\bottomrule
 	\end{tabular}
\end{adjustbox}
 \end{table*}

 Garcia, N. et al. \cite{garcia2017dress} proposed a framework of content-based fashion image retrieval from a visual stream. This model reduces the gap between users and fashion items displayed in videos. The model can easily relate the indexed database visual frames and clothing items. Target fashion products can easily be retrieved using these indexing techniques and frame retrieval with temporal aggregation from the visual stream. A dataset of 40 movies with more than 80 hours of duration is collected for experimental verification of the model. The query frame is captured with a webcam for processing and retrieval of fashion products using the proposed model. 
 Zhao, H. et al. \cite{zhao2021dress} proposed a Detect, Pick, and Retrieval Network (DPRNet) for retrieving fashion products from videos. The DPRNet automatically detects and picks the keyframes from the input visual stream without duplications to reduce the computational time and cost of the proposed model. A multi-task loss function is proposed to enhance the performance of retrieval fashion items from keyframes. A new large-scale dataset is proposed to evaluate the model's performance, which contains 818 videos and 21614 fashion product images.
 Zhang, Y. et al. \cite{zhang2021exploring} proposed a multimodal Visual-Audio Composition Alignment Network (VACANet) for CBFIR from the audio and visual stream. The audio-visual composition module is designed to discriminate the residual and attentive entities by knowing semantic embedding from audio and visual streams. After that, a domain constraint alignment technique for video-to-image adaptation and a quality alignment technique is designed for accurate fashion product retrieval. The audio-visual compositional and alignment of fashion products increase the accuracy of the desired fashion product. 
 Zhang, Y. Wang, Q. et al. \cite{zhang2021fashion} proposed a Fashion Focus framework for CBFIR using multiple modalities from the input visual stream. The main focus of the proposed framework is to localize the fashion item images from the online video. Various media, like visual content, audio, interaction context, and linguistic features, are jointly examined through the proposed Fashion Focus multimodal retrieval framework. The proposed framework directs users toward relevant fashion item exhibitions while watching online visual streams. 
 Godi, M. et al. \cite{godi2022movingfashion} proposed the MovingFashion standard dataset for CBFIR by tracking fashion items from multiple visual frames. The author presents a publicly available benchmark dataset for video-to-shop challenges. The dataset contains approximately 15 thousand videos from social platforms like TikTok, and Instagram, with their corresponding fashion product shop images. Additionally, the framework using SElf-Attention Multi-frame (SEAM) Match-RCNN architecture is also proposed that automatically associates visual frames from a given video with their corresponding shop fashion product image.

 \section{Content-Based Fashion Item Retrieval (CBFIR) datasets}
 \label{CBFIR-datasets}
 
 There are several different datasets that can be used with content-based fashion image retrieval techniques. These fashion-related databases are a major element in the accelerated advancement of multiple content-based fashion image retrieval methods. The datasets used in the different image-guided, image \& text-guided, sketch-guided, and video-guided CBFIR methods are briefly explained below. 
 
 \subsection{Image-guided CBFIR datasets}
 \label{image-datasets}
 
 Different datasets used for image-guided CBFIR tasks are explained herewith, and their comparative analysis are presented in Table \ref{tab:image datasets}.
 
 \begin{table*}[]
 	\centering
 	\footnotesize
 	\caption{Comparative analysis of different image-guided CBFIR datasets}
 	\label{tab:image datasets}
 	\begin{adjustbox}{width=1.1\textwidth, center=\textwidth}
 	\begin{tabular}{||p{3cm}||p{0.6cm}|p{1.2cm}|p{1.6cm}|p{1.5cm}|p{1.2cm}|p{1.8cm}|p{1.5cm}|}
 		\bottomrule
 		\rowcolor[HTML]{7CACF8} 
 		{\textbf{Dataset}} & {\textbf{Year}} & {\textbf{Total No. of images}} & {\textbf{No. of categories / styles}} & {\textbf{No. of attributes}} & {\textbf{Image-pairs}} & {\textbf{landmarks/ key-points}} & \textbf{Publicly available} \\
 		\toprule
 		{Plagiarized   Fashion} & 2020 & 60,000 & 4 & - & - & Yes & - \\
 		{RingFIR} & 2020 & 2,651 & 46 & - & - & - & Yes \\
 		{DeepFashion2} & 2019 & 491,000 & 13 & - & 873,000 & 801000   & Yes \\
 		{FashionAI} & 2019 & 357,000 & 6 & 245 & - & 24 & - \\
 		{FindFashion} & 2019 & 565,041 & - & 3 & 382,230 & b.box-Yes & Yes \\
 		{Shopping100k} & 2018 & 101,021 & 16 & 151 & - & - & - \\
 		{ModaNet} & 2018 & 55,176 & 13 & - & - & b.box-Yes & Yes \\
 		{FashionStyle14} & 2017 & 13,126 & 14 & - & - & - & Yes \\
 		{Polyvore-Outfit} & 2017 & 21,889 & 11 & - & 164,379 & - & Yes \\
 		{DeepFashion} & 2016 & \textgreater{}800,000 & 50 & 1000 & 300,000 & 4 $\sim$8 & Yes \\
 		{DARN} & 2015 & 540,000 & 20 & 179 & 91,390 & - & - \\
 		{Exact   Street2Shop} & 2015 & 425,040 & 11 & - & 39,479 & - & Yes \\
 		{CelebA} & 2015 & 202,599 & - & 40 & - & 5 & Yes \\
 		{UT Zappos50K} & 2014 & 50,025 & 4 & 8 & 11,363 & - & Yes \\
 		{Street2Shop} & 2012 & 25,025 & 4 & 15 & - & - & - \\
 		\bottomrule
 	\end{tabular}
\end{adjustbox}
 \end{table*}

 \textbf{DeepFashion:} dataset was proposed by Liu, Z. et al. \cite{liu2016deepfashion} in 2016. Many researchers use this dataset for image-guided CBFIR tasks. The dataset contains more than 800000 images from distinct scenarios like the consumer, stores, and street snapshots. These images are labelled according to 50 different categories and 1000 attributes. The DeepFashion dataset is publicly available for researchers. 
 \textbf{DeepFashion2:} dataset was proposed by Ge, Y. et al. \cite{ge2019deepfashion2} in 2019. This dataset contains 491000 images from different online shopping stores and consumers. All these images are labelled according to 13 popular categories of fashion images. DeepFashion2 contains a large number of shop-consumer image pairs, i.e., about 873000, which is massive compared to the DeepFashion dataset. \textbf{DARN:} dataset was proposed by Huang, J et al. \cite{huang2015cross} in 2015. This dataset contains 450,000 in-shop images and 90,000 consumer images. These images are labelled according to 20 different categories. Many researchers use the DARN dataset to evaluate fashion image retrieval models. 
 
 \textbf{Shopping100k:} dataset was proposed by Ak, KE. et al. \cite{ak2018efficient} in 2018. This dataset contains 101,021 fine-grained images. These images are labelled according to 16 different categories and 151 attributes, which are mainly categorized under general (gender, category, etc.) and special (collar, sleeve length, etc.) attributes. 
 \textbf{FashionStyle14:} dataset was proposed by Takagi, M. et al. \cite{takagi2017makes} in 2017. This dataset is proposed for the prediction of various fashion styles in the industry. This dataset contains 13,126 images, which are categorized into 14 different fashion styles. 
 \textbf{FashionAI:} dataset was proposed by Zou, X. et al. \cite{zou2019fashionai} in 2019. This large hierarchical dataset contains a total of 357,000 images from 6 different women's clothing categories. Twenty-four key points are used to label the different categories of women's clothing vastly. 
 
 \textbf{Street2Shop:} dataset was proposed by Liu, S. et al. \cite{liu2012street} in 2012. This dataset contains 25,025 images from online shopping and daily photo sources. Fifteen different attributes are labelled against each image with the perspective of the collected image's upper body and lower body. 
 \textbf{Exact Street2Shop:} dataset was proposed by Hadi Kiapour, M. et al. \cite{hadi2015buy} in 2015. This dataset contains a total of 425,040 images, where 404,683 images are collected from 25 different shops and 20,357 are collected from different unconditioned/uncontrolled street images. All these images are labelled according to 11 main categories of fashion items. 
 \textbf{UT Zappos50K:} dataset was proposed by Yu, A. and Grauman, K. \cite{yu2014fine} in 2014. This data set contains a total of 50,025 images of different shoes collected from Zappos.com. All these shoe images are divided into boots, slippers, sandals, and shoe categories. Additionally, each shoe image is labelled with eight different attributes (like materials, gender, etc.) to filter the shoes. 
 \textbf{Polyvore-Outfit:} dataset was proposed by Han, X. et al. \cite{han2017learning} in 2017. This dataset consists of 21,889 outfit images from the source website (polyvore.com), out of which 17,316, 1,497, and 3,076 images are used for train, val, and test splits, respectively. 
 
 \textbf{CelebA:} dataset was proposed by Liu, Z. et al. \cite{liu2015deep} in 2015. This dataset contains a total of 202,599 face images; almost ten thousand different face identities are collected, along with twenty images for each identity. Each image in the dataset is labelled with five key points and forty face attributes. 
 \textbf{RingFIR:} dataset was proposed by Islam, SM. et al. \cite{islam2021ringfir} in 2020. This dataset contains a total of 2,651 images from the golden earring catalogue. This dataset is designed to retrieve jewelry earring items from the given query image. 
 \textbf{Plagiarized fashion:} dataset was proposed by Lang, Y. et al. \cite{lang2020plagiarism} in 2020. This dataset consists of a total of 60,000 images of plagiarized clothes, out of which 40k images are used for training and 20k for testing. All these images are categorized into four different categories. 
 \textbf{ModaNet:} dataset was proposed by Zheng, S. et al. \cite{zheng2018modanet} in 2018. This dataset contains a total of 55,176 street images. All these images are divided into 13 different fashion categories with polygon annotations.  
 \textbf{FindFashion:} dataset was proposed by Kuang, Z. et al. \cite{kuang2019fashion} in 2019. This dataset contains a total of 565,041 images. All the images in this dataset are labelled with the perspective of cropping, occlusion, and view attributes to enhance the performance of retrieval tasks in the fashion industry.
 
 \subsection{Image \& text-guided CBFIR datasets}
 Different datasets used for image \& text-guided CBFIR tasks are explained herewith, and their comparative analysis are presented in Table \ref{tab:imageANDtext  datasets}.
 
 \begin{table*}[]
 	\caption{Comparative analysis of different image \& text-guided CBFIR datasets}
 	\footnotesize
 	\centering
 	\label{tab:imageANDtext  datasets}
 	\begin{adjustbox}{width=1.1\textwidth, center=\textwidth}
 	\begin{tabular}{||p{3cm}||p{0.6cm}|p{1.2cm}|p{1.6cm}|p{1.5cm}|p{1.2cm}|p{1.8cm}|p{1.5cm}|}
 		\bottomrule
 		\rowcolor[HTML]{7CACF8} 
 		{\textbf{Dataset}} & {\textbf{Year}} & {\textbf{Total No. of images}} & {\textbf{No. of categories / styles}} & {\textbf{No. of attributes}} & {\textbf{Image-pairs}} & {\textbf{Relative  captions}} & \textbf{Publicly available} \\
 		\bottomrule
 		{FashionIQ} & 2021 & 77,684 & 3 & 49,464 & Yes & 60,272 & Yes \\
 		{CIRR} & 2021 & 21,552 & - & - & 36,000 & - & Yes \\
 		{Amazon} & 2018 & 53,689 & 200 & - & 5000 & 5000 & No \\
 		{Fashion-Gen} & 2018 & 293,008 & 48 & - & 293,008 & 293,008 & Yes \\
 		{Fashion200K} & 2017 & 200,000 & 5 & 4404 & Yes & 5000 & Yes \\
 		{Street   Fashion Style} & 2017 & 293,105 & 43 & - & - & - & Yes \\
 		{E-commerce} & 2017 & 431,841 & 32 & - & Yes & - & No\\
 		\bottomrule
 	\end{tabular}
\end{adjustbox}
 \end{table*}

 DeepFashion \cite{liu2016deepfashion} and DARN \cite{huang2015cross} datasets are explained in section \ref{image-datasets} already; both of these datasets are also used in image \& text-guided CBFIR models. 
 \textbf{FashionIQ:} dataset was proposed by Wu, H. et al. \cite{wu2021fashion} in 2021. Many researchers use this dataset for image \& text-guided CBFIR tasks. The dataset contains a total of 77,684 images which are categorized into dresses, shirts, and tops\&tees categories. These images are labelled according to approximately 50k attributes and 60k relative captions. The FashionIQ dataset is publicly available for researchers.
 \textbf{Fashion200K:} dataset was proposed by Han, X. et al. \cite{han2017automatic} in 2017. This dataset contains around 200,000 fashion product images with descriptions of more than four words from different online shopping platforms. All the images are divided into five categories with long descriptions. The fashion200k dataset is also available online for researchers.  
 \textbf{CIRR:} dataset was proposed by Liu, Z. et al. \cite{liu2021image} in 2021. The CIRR dataset is designed specifically for fashion image retrieval tasks and contains real-life images extracted from a famous natural language reasoning database, NLVR2. 
 \textbf{Amazon:} dataset was proposed by Liao, L. et al. \cite{liao2018interpretable} in 2018. This dataset contains a total of 53,689 images of fashion products from online amazon web databases. All these images are divided into 200 different clothing categories and contain around 5000 images and textual description pairs. 
 
 \textbf{Fashion-Gen:} dataset was proposed by Rostamzadeh, N. et al. \cite{rostamzadeh2018fashion} in 2018. This dataset contains a total of 293,008 fashion product high-resolution images. These images are divided into 48 different fashion product categories, and each image is labelled with its relative textual description. 
 \textbf{Street Fashion Style:} dataset was proposed by Gu, X. et al. \cite{gu2017understanding} in 2017. The Street Fashion Style dataset contains a total of 293,105 outfit images from Chictopia. All these images are divided into four categories for season classification, 15 categories for style classification, and 24 categories for garments classification.  
 \textbf{E-commerce:} dataset was proposed by Rubio, A. et al. \cite{rubio2017multi} in 2017. This dataset contains a total of 431,741 images from various fashion e-commerce databases. In this dataset collection, all the images are divided into 32 different fashion item categories with corresponding textual description pairs. 
 
 \subsection{Sketch-guided CBFIR datasets}
 
 Different datasets used for the sketch-guided CBFIR tasks are explained herewith, and their comparative analysis are presented in Table \ref{tab:sketch-datasets}.
 
 \begin{table*}[]
 	\centering
 	\footnotesize
 	\caption{Comparative analysis of different sketch-guided CBFIR datasets}
 	\label{tab:sketch-datasets}
 	\begin{adjustbox}{width=1.1\textwidth, center=\textwidth}
 	\begin{tabular}{||p{3cm}||p{0.6cm}|p{1.2cm}|p{1.6cm}|p{1.5cm}|p{1.2cm}|p{1.8cm}|p{1.5cm}|}
 		\bottomrule
 		\rowcolor[HTML]{7CACF8} 
 		{\textbf{Dataset}} & {\textbf{Year}} & {\textbf{Total No. of   images}} & {\textbf{No. of   categories}} & {\textbf{Types/ styles}} & {\textbf{Sketch-photo pairs}} & {\textbf{No. of attributes}} & \textbf{Publicly available} \\
 		\bottomrule
 		{ Fashion Image} & 2021 & 36,074 & 4 & 26 & 36,074 & - & No \\
 		{QMUL-shoes} & 2016 & 1,432 & 1 & 5 & 419 & 21 & Yes \\
 		{QMUL-chairs} & 2016 & 1,432 & 1 & 4 & 297 & 15 & Yes \\
 		{TU Berlin} & 2012 & 20,000 & 250 & - & 20,000 & - & No \\
 		\bottomrule
 	\end{tabular}
\end{adjustbox}
 \end{table*}

 \textbf{Fashion Image:} dataset was proposed by Lei, H. et al. \cite{lei2021new} in 2021. Researchers used this dataset for sketch-guided CBFIR tasks. The dataset contains a total of 36,074 images categorized into clothes, skirts, pants, and shoes. The cloth category is further divided into 11 types, pants into four types, skirts into six types, and shoes into five further types. 
 \textbf{QMUL-shoes:} dataset was proposed by Yu, Q. et al. \cite{yu2016sketch} in 2016. The dataset contains a total of 1,432 images and sketches categorized into the shoe category. Rich annotations are provided by fashion experts using finger-sketch-based sketch-photo pairs to evaluate the performance of sketch-based CBFIR methods. Five types of shoes, i.e., boots, informal, formal, high-heals, and ballerinas images, are collected from UT-Zap50K. Four hundred nineteen sketch-photo pairs are used for fashion item retrieval. 
 \textbf{QMUL-chairs:} dataset was proposed by Yu, Q. et al. \cite{yu2016sketch} in 2016. The dataset contains a total of 1,432 images and sketches categorized into chair categories. Four types of chairs, i.e., kid chairs, office chairs, couches, and desk chair images, are collected from Taobao, Amazon, and IKEA online shopping websites. Two hundred ninety-seven sketch-photo pairs are used for fashion item retrieval.
 \textbf{TU Berlin:} dataset was proposed by Eitz, M. et al. \cite{eitz2012humans} in 2012. This dataset contains a total of 20,000 sketches drawn by hand sketching, which are categorized further into 250 different categories. 
 
 \subsection{Video-guided CBFIR datasets}
 
 Different datasets used for video-guided CBFIR tasks are explained herewith, and their comparative analysis are presented in Table \ref{tab:Video-datasets}.
 
 \begin{table*}[]
 	\centering
 	\footnotesize
 	\caption{Comparative analysis of different video-guided CBFIR datasets}
 	\label{tab:Video-datasets}
 	\begin{adjustbox}{width=1.1\textwidth, center=\textwidth}
 	\begin{tabular}{||p{2.4cm}||p{0.6cm}|p{1.2cm}|p{1.6cm}|p{1.5cm}|p{1.2cm}|p{1.8cm}|p{1.5cm}|}
 		\bottomrule
 		\rowcolor[HTML]{7CACF8} 
 		{\textbf{Dataset}} & {\textbf{Year}} & {\textbf{Total No. of clips/videos}} & {\textbf{No. of   categories}} & {\textbf{Label annotation}} & {\textbf{Frame per second}} & {\textbf{No. of categories}} & \textbf{Publicly available} \\
 		\bottomrule
 		{MovingFashion} & 2022 & 14855 & Social media & Yes & 30 & - & Yes \\
 		{Video-to-shop} & 2021 & 818 & e-commerce & - & - & - & No \\
 		{Tmall Fashion Retrieval} & 2021 & 3,123 & Taobao & Yes &  & 14 & No\\
 		\bottomrule
 	\end{tabular}
\end{adjustbox}
 \end{table*}
 
 DeepFashion \cite{liu2016deepfashion} and DeepFashion2 \cite{ge2019deepfashion2}, already explained in the image-guided CBFIR datasets section \ref{image-datasets}, are also used for video-guidance CBFIR tasks. 
 \textbf{MovingFashion:} dataset was proposed by Godi, M. et al. \cite{godi2022movingfashion} in 2022. Researchers used this dataset for video-guided CBFIR tasks. This dataset contains a total of 14855 video clips from Net-A-Porter (e-commerce website) and different social media platforms like TikTok, Instagram, etc. Each video contains 30 frames per second and a total of 5.85 million annotations for these frames. The MovingFashion dataset is also available online for researchers.  
 \textbf{Video-to-Shop:} dataset was proposed by Zhao, H. et al. \cite{zhao2021dress} in 2021. This dataset contains a total of 818 video clips from well-known e-commerce Tmall and Taobao online platforms used for fashion retrieval tasks.  
 \textbf{Tmall Fashion Retrieval:} dataset was proposed by Zhang, Y. et al. \cite{zhang2021exploring} in 2021. This dataset contains a total of 3,123 short video clips from well-known e-commerce Tmall and Taobao online platforms used for fashion retrieval tasks. The clothing from these videos is categorized into the bags, snacks, shirt, digital, furniture, pants, cosmetics, shoes, toys, beverages, underdress, dress, accessories, and other categories.

 \section{Discussion and Comparative analysis}
 \label{Discussion-comparativeAnalysis}
 
 Any retrieval mechanism must allow consumers to accurately and quickly specify their search requirements. In conventional fashion retrieval models, a sample text-based query interface is presented to the user. The text-based mechanism does not work well with a content-based fashion image retrieval system because it uses visual queries to express a user's needs. The easiest and most enticing approach is to provide a reference image as a query, where the user provides a reference image to express their desired requirements about fashion products or items to launch the retrieval. Following is the list of some key challenges of CBFIR models. 
 
 \begin{itemize}
 	\item Content-based fashion image retrieval must process high-dimensional and contiguous data like color histograms. For speedy and effective searching from a collection of images, finding index structures is still an unresolved issue.
 	\item The major semantic gap between low-level visual signals and high-level semantic indicators (such as sleeve length and neckline) that decode users' query intent presents a significant barrier to CBFIR approaches.
 	\item Multiple media, like images and text, is a prominent part of content-based fashion image retrieval. Due to the fact that each media is located in its own distinct feature space, there is a feature gap between them.
 	\item The raw data in the fashion industry frequently originates from several fashion domains, including product, runway, and street fashion. It is concluded from the literature that fashion product shots, street fashion shots, and designer runway shots all have distinct visual features and accompanying textual information. That's why fashion products from different domains generally reflect different semantic meanings. Consequently, there are significant differences in clothing data from various fashion domains.
 \end{itemize}
 
 Limitations and issues of various content-based fashion image retrieval methods mentioned in section 3 are highlighted below. Possible future research directions of image-guided CBFIR, image \& text guided CBFIR, sketch-guided CBFIR, and video-guided CBFIR are discussed separately.
 
 \subsection{Image-guided CBFIR}
 
 The use of imagery for fashion items or product searches is one of the most often utilized techniques since visual information is easier to access and contains more data than text-based information does. To achieve the objective of image-guided CBFIR, the retrieval systems face some key issues. The major issue is there are too many components in the apparel images, including texture, color, style, and shape. But, all of these elements are hardly ever covered by general retrieval techniques. DeepFashion and DeepFashion2 are the most widely used datasets in the field of content-based fashion image retrieval. These two large clothing datasets will be extremely useful in future content-based fashion image retrieval research. The dress landmark recognition technique can also handle dress details, occlusion, and deformation. The accuracy of fashion image retrieval is greatly improved using this landmark recognition technique; however, a large amount of labelling and annotation information is required, as is professional knowledge of the clothing industry. Key limitations of image-guided CBFIR models are discussed below with their possible future research directions.

 \textbf{Attribute labelling:} The majority of image-guided CBFIR models with attribute guidance rely on dress attribute labels for support. It necessitates extensive clothing labelling and is a labour-intensive and time-consuming task. More research is needed to determine how to minimize clothing attribute labelling, improve retrieval accuracy, and save time and resources.
 
 \textbf{Style-aware fashion image retrieval:} A retrieved fashion image may or may not suits the occasion or style of event to be attended. In recent years, only a few authors have worked to retrieve style-aware fashion items. Therefore, content-based fashion image retrieval with style guidance is a future research direction for researchers.

 \subsection{Image \& Text-guided CBFIR}
 
 Image \& text-guided CBFIR models have a huge impact on retrieving desired fashion items or products in the fashion industry. The majority of the known embedding techniques only take the "heterogeneous similarity" requirement into account. When using the image and text-guided CBFIR, heterogeneous similarity forces conceptually related (text, image) pairs to be near one another in the commonly hidden space to acquire cross-modal similarity with the training (text, image) pairs. The similarity among several views-points can thus be measured using the learnt metric distance. To improve the efficiency of cross-modal fashion image retrieval systems for multi-domain CBFIR tasks, homogeneous and heterogeneous similarity constraints must be retained. Key limitations of image \& text-guided CBFIR models are discussed below with their possible future research directions.
 
 \textbf{Region-specific clothing:} In the image \& text-guided content-based fashion image retrieval task, there is a lack of region-specific attribute labelling of clothing items. It is a challenging and time-consuming task to label each region of clothing item, but it improves the customers' satisfaction with the retrieved fashion item. Currently, no public database is available that contains region-specific descriptions/annotations of clothing items. Therefore, it is a possible future research direction for the researchers to increase the retrieval accuracy of the image \& text-guided CBFIR. 
 
 \textbf{Retrieval with customer's multi-turn textual feedback:} Multi-turn textual feedback is a good idea to retrieve the desired target image, but extracting specific clothing attributes from a customer's diversified conversational textual feedback is very challenging. Multi-turn textual feedback for fashion image retrieval needs to be optimized in the future by mentioning some specific clothing item attributes like color, style, shape, pattern, texture, fabric, collar, length, etc.

 \subsection{Sketch-guided CBFIR}
 
 The query approach for the image-guided and video-guided content-based fashion image retrieval method is constrained for optimum usage because it is likely that the user lacks an appropriate reference video clip or image. Thus, a more user-friendly and efficient method is needed to assist these consumers in finding fashion items or products that correspond to their requirements and preferences. With the quick advancement of visual sensor innovations, modern-day tablets, mobile phones, drones, and smart cameras are just a few examples of innovative mobile sensors. The particular mobile sensor technology of tablets and smartphones offers a user-friendly "human-computer interface" that allows most users (even without much usage experience) to simply manipulate the mobile sensors by touching and sketching diagrams on the screen. Sketch-guided is a powerful method of human-machine communication that offers a user-friendly and natural way to search through a vast collection of fashion product images with the help of reference sketch images.
 Sketch-guided CBFIR technology is in progress, and still, there is a lot of room available for researchers to improve retrieval accuracy using sketching techniques. As sketch-guided fashion image retrieval is an effective yet fast method for retrieving desired clothes from large image repositories, this technology is still facing some key challenges. Like, 
 \begin{itemize}
 	\item Representing clothes with sketches is fundamentally ambiguous because a sketch is a rough illustration that only contains partial information about a clothes image.
 	\item Sketches are instinctually devoid of visual cues (for example, the absence of texture and colour information), making the implementation of conventional image-oriented schemes difficult.
 	\item Moreover, one cloth image can be sketched in different styles by different users, which can affect the retrieval accuracy using this sketch-guided CBFIR method.
 \end{itemize}

 Key limitations of sketch-guided CBFIR models are discussed below with their possible future research directions.
 
 \textbf{Clothing component segmentation:} Clothing component segmentation plays a vital role in improving retrieval accuracy using the sketch-guided fashion image retrieval technique. Parallel processing of clothing component segmentation along with customer feedback incorporation is a possible future research direction in this CBFIR field.  
 
 \textbf{Consumer purchase patterns:} Design and implement a system that incorporates the real-time consumer purchase patterns from big shopping malls into the system to increase the desired product retrieval accuracy. 
 
 \textbf{Sketch-based fashion item datasets:} As sketch-guided content-based fashion image retrieval techniques are a hot topic these days. Therefore, a huge annotated collection of different fashion categories are still required to validate the accuracy of sketch-based fashion item retrieval.

 \subsection{Video-guided CBFIR}
 
 items or products from a huge collection of fashion products. As "consumer-to-shop" fashion image retrieval has advanced significantly,  which locates identical clothes in online shops using images from customers' reference queries. But few studies still exist on locating desired fashion images from visual streams. Video-guided CBFIR methods are more challenging than consumer-to-shop because of the different viewpoints, crops, and occlusion for videos that feature apparel detection. The aforementioned variances of the same outfit could lead to a large visual disparity and unquestionably significantly decrease the video-guided CBFIR performance. Recognizing all fashion item proposals in the visual stream and accurate classification of these proposals into instance levels is another problem. Due to different presentation viewpoints and the disruption of background-foreground items, live-stream localization is more difficult than typical video-to-shop fashion image retrieval. The problem of inadequate data still affects the performance of live-steam fashion item retrieval. However, the addition of audio clues and user responses in the form of text with live-steam video increases the contextual details of the stream. Some possible future research directions for video-guided CBFIR models are discussed below. 
 
 \textbf{Insufficient data to deal with live-stream videos:} Generally, live-stream video content is guided by interaction comments and audio interpretations, which supplement rich contextual data with sparse annotations. In content-based fashion image retrieval, still, there are inadequate databases available to deal with live-streams videos for retrieval of desired fashion item.
 
 \textbf{Inadequate video-to-shop datasets:} Video-to-shop content-based fashion image retrieval is a relatively new research area these days. Not many video-to-shop datasets are available publicly for researchers to validate video-guided CBFIR models. Therefore, in the future, more video-to-shop datasets need to be designed to increase the accuracy of video-to-shop content-based fashion image retrieval tasks.

 \section{Conclusion}
 \label{concl}
 
 In this survey, we discussed the recent methods and advancements of content-based fashion image retrieval. Content-based fashion image retrieval (CBFIR) has been widely used in our daily life for searching fashion images or items from online platforms. Various CBFIR methods are categorized into image guidance, image \& text guidance, sketch guidance, and video guidance retrieval methods. This article reviews the recent up-to-date methods and advancements in fashion image retrieval tasks. A novel fine-grained taxonomy of numerous CBFIR methods is designed. A detailed comparative analysis is provided for each group of CBFIR methods by discussing the FIR network architectures, loss functions, datasets, and evaluation metrics. A brief discussion on image-guided, image \& text-guided, sketch-guided, and video-guided CBFIR datasets is also provided separately. Comparative analysis of different datasets used in image-guided, image \& text-guided, sketch-guided, and video-guided CBFIR methods is also provided. Finally, the key challenges and limitations of each fashion image retrieval method are discussed separately, along with their possible reseach directions. We hopefully anticipate that our proposed fine-grained taxonomy and detailed future directions in each category will promote future studies in content-based fashion image retrieval and helps researchers to better understand unresolved issues in this field.




%

%

\bibliographystyle{ACM-Reference-Format}
\bibliography{ref}

%
\appendix

\end{document}